\documentclass[11pt]{article}
\usepackage{jcappub}
\usepackage{graphicx}
\usepackage{natbib}
\usepackage{amsmath}
\usepackage{color,colordvi}
\usepackage{booktabs, multirow}
\usepackage[mediumspace,Gray,squaren]{SIunits}
\usepackage{hyperref}
\usepackage{aas_macros}
\usepackage[utf8]{inputenc}
\usepackage{amssymb,amsfonts}
\definecolor{darkblue}{rgb}{0,0,0.3}
\definecolor{darkgreen}{rgb}{0,0.3,0}
\hypersetup{
  pdftitle = {Planck x ROSAT},
  pdfauthor = {Hajian, Battaglia, et al.},
  colorlinks,%
  citecolor=darkblue,%
  filecolor=black,%
  linkcolor=blue,%
  urlcolor=blue
}

\newcommand{\head}[1]{\textnormal{\textbf{#1}}}

\addunit{\jansky}{\ensuremath{\mathrm{Jy}}}
\addunit{\degword}{\ensuremath{\mathrm{deg}}}
\newcommand{\be}{\begin{equation}}
\newcommand{\ee}{\end{equation}}
\newcommand{\bea}{\begin{eqnarray}}
\newcommand{\eea}{\end{eqnarray}}
\newcommand{\rmn}{\mathrm}
\newcommand{\dd}{\mathrm{d}}

\begin{document}

\title{Measuring the Thermal~Sunyaev-Zel'dovich~Effect Through the Cross~Correlation of Planck and WMAP Maps with ROSAT Galaxy Cluster Catalogs}

\author[a]{Amir Hajian,}
\emailAdd{ahajian@cita.utoronto.ca}
\author[b]{Nicholas Battaglia,}
\emailAdd{nbattaglia@cmu.edu}
\author[c]{David N. Spergel,}
\author[a]{J.~Richard~Bond,}
\author[d]{Christoph Pfrommer,}
\author[e,f]{Jonathan L. Sievers}

\affiliation[a]{Canadian Institute for Theoretical Astrophysics, University of Toronto,\hspace{1em}Toronto, ON M5S~3H8, Canada}
\affiliation[b]{McWilliams Center for Cosmology, Wean Hall, Carnegie Mellon University, 5000 Forbes Ave., Pittsburgh, PA 15213, USA}
\affiliation[c]{Department of Astrophysical Sciences, Peyton Hall, Princeton University, Princeton, NJ 08544, USA}
\affiliation[d]{Heidelberg Institute for Theoretical Studies, Schloss-Wolfsbrunnenweg 35, D-69118 Heidelberg, Germany}
\affiliation[e]{Astrophysics and Cosmology Research Unit, University of Kwazulu-Natal, Westville, Durban 4000, South Africa}
\affiliation[f]{Joseph Henry Laboratories of Physics, Jadwin Hall, Princeton University, Princeton NJ 08544, USA}

\abstract{We measure a significant correlation between the thermal Sunyaev-Zel'dovich effect in the Planck and WMAP maps and an X-ray cluster map based on ROSAT. We use the 100, 143 and 343 GHz Planck maps and the WMAP 94 GHz map to obtain this cluster cross spectrum. We check our measurements for contamination from dusty galaxies using the cross correlations with the 220, 545 and 843 GHz maps from Planck. Our measurement yields a direct characterization of the cluster power spectrum over a wide range of angular scales that is consistent with large cosmological simulations. The amplitude of this signal depends on cosmological parameters that determine the growth of structure ($\sigma_8$ and $\Omega_\rmn{M}$) and scales as $\sigma_8^{7.4}$ and $\Omega_\rmn{M}^{1.9}$ around the multipole $(\ell) \sim 1000$. We constrain $\sigma_8$ and $\Omega_\rmn{M}$ from the cross-power spectrum  to be  $\sigma_8  (\Omega_\rmn{M}/0.30)^{0.26} = 0.8 \pm 0.02$. Since this cross spectrum produces a tight constraint in the $\sigma_8$ and $\Omega_\rmn{M}$ plane the errors on a $\sigma_8$ constraint will be mostly limited by the uncertainties from external constraints. Future cluster catalogs, like those from eRosita and LSST, and pointed multi-wavelength observations of clusters will improve the constraining power of this cross spectrum measurement. In principle this analysis can be extended beyond $\sigma_8$ and $\Omega_\rmn{M}$ to constrain dark energy or the sum of the neutrino masses.}

\maketitle
\flushbottom

\section{Introduction}

Galaxy clusters are the rarest collapsed objects in the Universe and sign posts for structure formation. Their abundances are extremely sensitive to the underlying cosmological parameters, such as the amplitude of the matter power spectrum. Current cluster surveys \citep[e.g.,][]{Vand2010,Sehg2011,Planck-XX} or those just beginning \citep[e.g.,][]{DES,eRose} aim to take advantage of this sensitivity and produce precise cosmological constraints from their growth of structure measurements. However, there is a persistent and unresolved issue when using clusters for precision cosmological measurements which is the large uncertainties associated with connecting observable cluster quantities with its mass (i.e., the observable-mass relation) \citep[e.g.,][]{Benson/2011,Hass2013,Rozo2013}. This issue was recently highlighted when Planck cluster cosmology constraints were shown to be in tension with their primary Cosmic Microwave Background (CMB) constraints \citep{Planck-XX}.

Clusters are either observed through the galaxies they contain or the intracluster medium (ICM), which is a hot ($\gtrsim 10^7$ K) plasma. Both the galaxies and the ICM have properties that correlate with cluster mass. For galaxies these properties are number of galaxies within a cluster or the {\it richness} \citep[e.g.,][]{Abel1989,Glad2005,Kost2007,Rykf2013} that correlates with its mass. For the ICM it is the thermodynamic properties of the ICM, such as temperature or X-ray luminosity measured by bremsstrahlung emission that correlates with cluster mass. The X-ray luminosity is proportional to the density squared, which makes it sensitive to the complex inner regions of clusters and difficult to model correctly. Therefore, X-ray observables from cluster to cluster vary greatly depending on the state of the core region. Furthermore, flux limited X-ray samples preferential select relaxed cool-core clusters.

An alternative and complementary probe of the ICM is the thermal Sunyaev-Zeldovich (tSZ) effect \citep{Sunyaev:1972}: as the CMB photons inverse-Compton scatter off of the hot electrons in the ICM, their blackbody spectrum is distorted. The tSZ effect is proportional to electron pressure along the line of sight and traces the total thermal energy in clusters. Pressure is a more stable thermodynamic quantity for clusters than the density. There are significant contributions to the tSZ flux that come from beyond the virial radius and it is less sensitive to the state of cluster core and selects clusters independent of their dynamical state (and redshift). Thus, modeling the tSZ is easier than the X-ray signal.

Previously, statistical measurements of the SZ effect were made by stacking the Wilkinson Microwave Anisotropy Probe (WMAP) temperature maps on locations of clusters. These measurements stacked on X-ray cluster catalogs \citep[e.g.,][]{Afsd2005,Lieu2006,Afsd2007,Atri2008,Dieg2010,Melin2011,WMAP7cos,Siavash}, optical cluster catalogs, \citep[e.g.,][]{Biel2007}, or both \citep[e.g.,][]{HM2004a}. In addition, other statistical detections were made cross correlating with X-ray cluster catalogs \citep{Benn2003,Hins2007} and galaxy surveys \citep{Fosa2003,Myers2004,HM2004b}. All of these measurements constrained average ICM quantities for the given sample selections, such as the average cluster profile. More recently, the Planck collaboration has performed several stacking analyses with the Planck data \citep{Planck10,PlanckeXI,PlanckeXII,PlanckIII,PlanckV,Planck11} and used these relationships in their cosmological analysis with clusters \citep{Planck-XX}. Beyond WMAP and Planck, measurements made by higher resolution CMB experiments such as the Atacama Cosmology Telescope (ACT) and the South Pole Telescope (SPT) as well as optimized experiments for SZ cluster detection such as Bolocam, the Arcminute Microkelvin Imager (AMI), MUSTANG and the Sunyaev Zel'dovich Array (SZA) have provided large amounts of quality SZ measurements of individual clusters \citep[e.g.,][]{TMarr2011,Benson/2011,TMarr2011,Will2011,Planck10,Marr2012,Sifo2013}, in addition to making average ICM measurements \citep[e.g.,][]{Plag2010,Bona2012,Saye2013}. Combining these large samples with mass proxies allows one to construct SZ-mass scaling relations that can be used for cluster cosmology.

The cluster observable-mass relation is commonly quantified through averaging over larger cluster samples as mentioned above. The result is a power power-law relation fit through the observable-mass plane, where the cluster masses are determined through weak lensing measurements or other previously {\it calibrated} mass proxies. In such analyses it is difficult to propagate the errors and account for biases and correlations between the observables, proxies and selection functions. Thus, assigning realistic error bars and uncertainties to cosmological measurements that rely on cluster masses is non-trivial.  

Measuring the spatial distribution of clusters, for example the angular correlation or power spectrum, is almost independent of this {\it error-prone} observable to mass conversion \citep[see e.g.,][]{KS2002}. The uncertainties in this approach come from modeling the signal for a given cluster and the selection function of the clusters. There is a large literature on the tSZ power spectrum and its prospects for cluster cosmology \citep[e.g.,][]{KS2002,Batt2010,Shaw2010,Trac2011}. In practice constraints from the tSZ power spectrum are limited by the modeling uncertainties mentioned above and the ability to separate its signal from the contributions of other secondary sources, like infrared point sources \citep[e.g.,][]{Shir2011,Dunk2011,Reic2012,Siev2013}.

In this paper we explore the cross correlation between a flux limited X-ray selected sample of clusters and all-sky CMB maps  from WMAP (94 GHz) and Planck (100, 143, 217, 353, 545 and 875 GHZ). This cross correlation is based on measuring the cross-power spectra and alleviates the effect of the correlated noisy modes that contaminate real-space cross-correlations. A somewhat similar analysis was attempted on WMAP 1-year data by \citep{Dieg2003} where no detectable signal was found. Here we take advantage of a cleaner X-ray data by using the overdensity map instead of the raw and contaminated X-ray map. This approach can be trivially generalized to any well characterized cluster catalog. Also using Planck data improves the analysis because of the higher sensitivity and resolution of Planck compared to WMAP. We demonstrate that even WMAP 9-year data show a significant cross-spectrum with the X-ray cluster overdensity map as expected. 

This paper is organized as follows: 
Section \ref{sec:data} describes the data and catalogs.
Section \ref{sec:meth} presents the methodology. We describe the analytical and numerical predictions in Section \ref{sec:thry}. The results are shown in Section \ref{sec:results}. We discuss these results and conclude in Section \ref{sec:concl}. Throughout this paper we use a $\Lambda$CDM cosmology consistent with the WMAP9 parameters \citep{WMAP9cos}, except where we explicitly state otherwise.

\section{Data} \label{sec:data}
\subsection{Planck Data}
We use six Planck maps from Planck's first data release \citep[][Planck Collaboration I]{Planck-I} at 100, 143, 217, 353, 545 and 875 GHZ in this analysis. The first two and the fourth maps are used to measure the signal while the remaining maps are used as null tests. The maps are at HEALPix \footnote{\url{http://healpix.jpl.nasa.gov}}  resolution 11 ($N_{\rm side}=2048$). And they are largely signal dominated in the range of multipoles we use in our analysis. The effective beam sizes for these maps are 9.651, 7.248, 4.990 arcmin but the Planck beams are asymmetric. This asymmetry couples the scan strategy of Planck into the beam window functions and makes the beam deviate from a Gaussian specially at small scales \citep[][Planck Collaboration VII]{Planck-VII}. We use the Reduced Instrument Model (RIMO) public beams of Planck as effective beams for each map.
These maps are not processed for foreground contamination and we use them in their raw format.

We use the published Planck + WMAP low-$\ell$ polarization + high-$\ell$ chains \citep{Planck16} for $\Lambda$CDM from the Planck Legacy Archive webpage\footnote{\url{http://pla.esac.esa.int/pla/aio/planckProducts.html}} when we derive the $\sigma_8 - \Omega_\rmn{M}$ contours.

\subsection{WMAP Data}
We use co-added inverse-noise weighted data from 36 single-year maps observed by WMAP at $94$ GHz (W-band, 4 channels and 9 years).
The maps are foreground cleaned (using the foreground template model discussed in \cite{hinshaw/etal:2007}) and are at HEALPix resolution 10 ($N_{\rm side}=1024$). The WMAP data are signal dominated on large scales \citep{Larson/etal:2011} and the detector noise dominates at smaller scales. The noise in WMAP data is a non-uniform (anisotropic) white noise that varies from pixel to pixel in the map. Pixel noise in each map is determined by $N_\rmn{obs}$ with the expression
\be \label{eq:noise}
\sigma = \sigma_0/\sqrt{N_\rmn{obs}},
\ee
where $\sigma_0$ is the noise for each differencing assembly and can be found on the WMAP data products webpage on LAMBDA\footnote{\url{http://lambda.gsfc.nasa.gov/product/map/dr5/m_products.cfm}}. $N_\rmn{obs}$ is the number of observations at each map pixel: regions with larger number of observations have lower noise variances and are given larger statistical weight. $N_\rmn{obs}$ is included in the maps available from the LAMBDA website. 

In all of our analysis we use pixel masks to exclude foreground-contaminated regions of the sky from the analysis. We use a galactic mask provided by Planck that masks 20\% of the sky around the galactic plane. We do not mask the point sources.

We use the published WMAP9 + SPT + ACT chains \citep{WMAP9cos} for $\Lambda$CDM on LAMBDA when we derive the $\sigma_8 - \Omega_\rmn{M}$ contours. 

\begin{figure}[t!]
\centering
\includegraphics[scale=0.6]{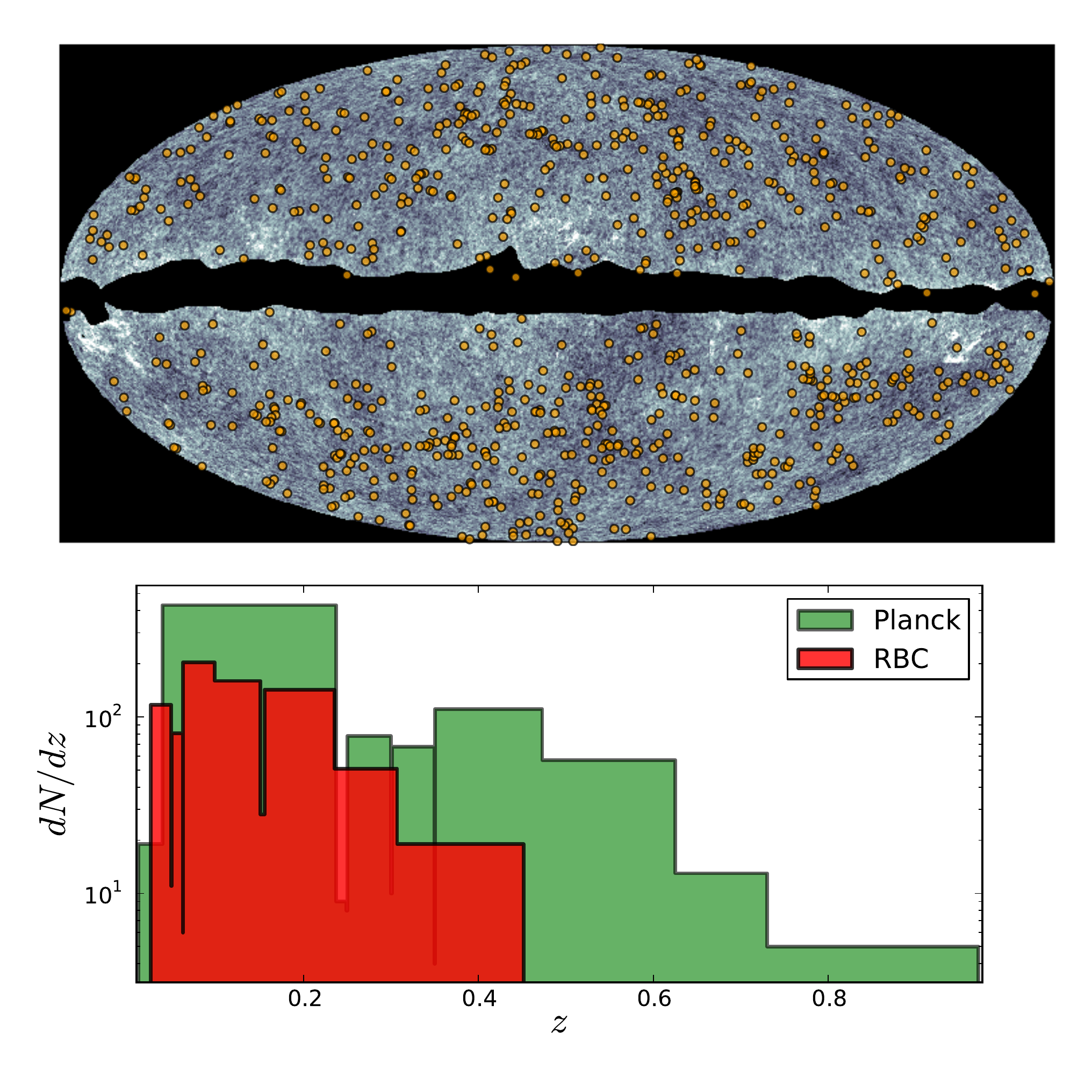}
\caption{\textit{Top:} Planck 100 GHz map and the cluster positions used in this analysis. The region in the center of the map is masked by the galaxy mask. \textit{Bottom:} redshift distribution of the X-ray cluster sample used in this analysis. The redshift distribution of the resolved Planck clusters from Planck's first year data release is also shown (although not used in this paper) for the sake of comparison.}
\label{fig:Map}
\end{figure}

\subsection{Cluster Catalog} \label{sec:clsl}
The catalog used in this analysis is a sub-sample of the MCXC X-ray cluster catalog \citep{piff2011}, which combines the REFLEX\citep{bohr2004}, BCS\citep{ebel1998,ebel2000} and CIZA\citep{ebel2002,koce2007} flux limited catalogs (hereafter the we will refer to this catalog as the RBC catalog) using the RoSAT all sky survey \citep[RASS;][]{voge1999}. The RBC catalog has a flux limit of $\sim 3\times 10^{-12}/ \rmn{erg}\,\rmn{s}^{-1},\rmn{cm}^{-2}$ and we assume that it is complete above this flux. We cut the catalog at $z = 0.04$ to reduce the cosmic variance from low redshift rare objects and alleviate the complications associated with projecting simulations below this redshift. The clusters masses were taken from the MCXC catalog, which uses a $L_\rmn{x}-M$ scaling relation from the REXCESS cluster sample \citep{prat2009} with an updated mass proxy calibration \citep{arnd2010}. The scaling relation is

\be \label{eq:lxm}
h(z)^{-7/3} \left(\frac{L_{500}}{10^{44}\rmn{erg}\,\rmn{s}^{-1}}\right) = C\left(\frac{M_{500}}{2 \times 10^{14} M_{\odot}}\right)^{1.64}
\ee

\noindent where $h(z) \equiv [\Omega_\rmn{M}(1+z)^3 + \Omega_{\Lambda}]^{1/2}$ and log$(C) = 0.274$. The scatter in the X-ray luminosity-mass relationship is $\pm$38 \%\citep{prat2009}, which implies a $\pm 24$\% scatter in the mass-X-ray luminosity relationship. The scatter is this larger, since we are using the X-ray luminosity without excising the core region. In our analysis, we incorporate the effect of the resulting Eddington bias on the sample through a Monte Carlo method that draws from the masses in our simulations (details in Sec. \ref{sec:num}).

We account for an additional bias of the $L_\rmn{x}-M$ relation since the cluster masses that enter this relation have been derived by assuming hydrostatic equilibrium. Simulations have shown that mass proxies that assume hydrostatic equilibrium are biased low between 10\% and 30\% \citep[e.g.,][]{Rasia2006,Lau2009,Batt2012a,Rasia2012} because of non-thermal contributions to the pressure support. When constructing our RBC catalog we applied an upward correction of 20\% to the MCXC catalog masses. We explore the uncertainty associated with this 20\% hydrostatic mass through varying our correction by $\pm 10$\% (see Sec \ref{sec:ThryUn}).

We show the spatial and redshift distributions of our cluster sample in Figure \ref{fig:Map}. Also plotted for comparison is the redshift distribution of the resolved Planck clusters. The histograms are done using the \textit{Bayesian Blocks} framework of \cite{BayesianBlocks} which chooses the bin sizes adaptively using the data\footnote{We use the astroML implementation \cite{astroML} of the Bayesian Blocks.}.

\begin{figure}[t!]
\centering
\includegraphics[scale=0.6]{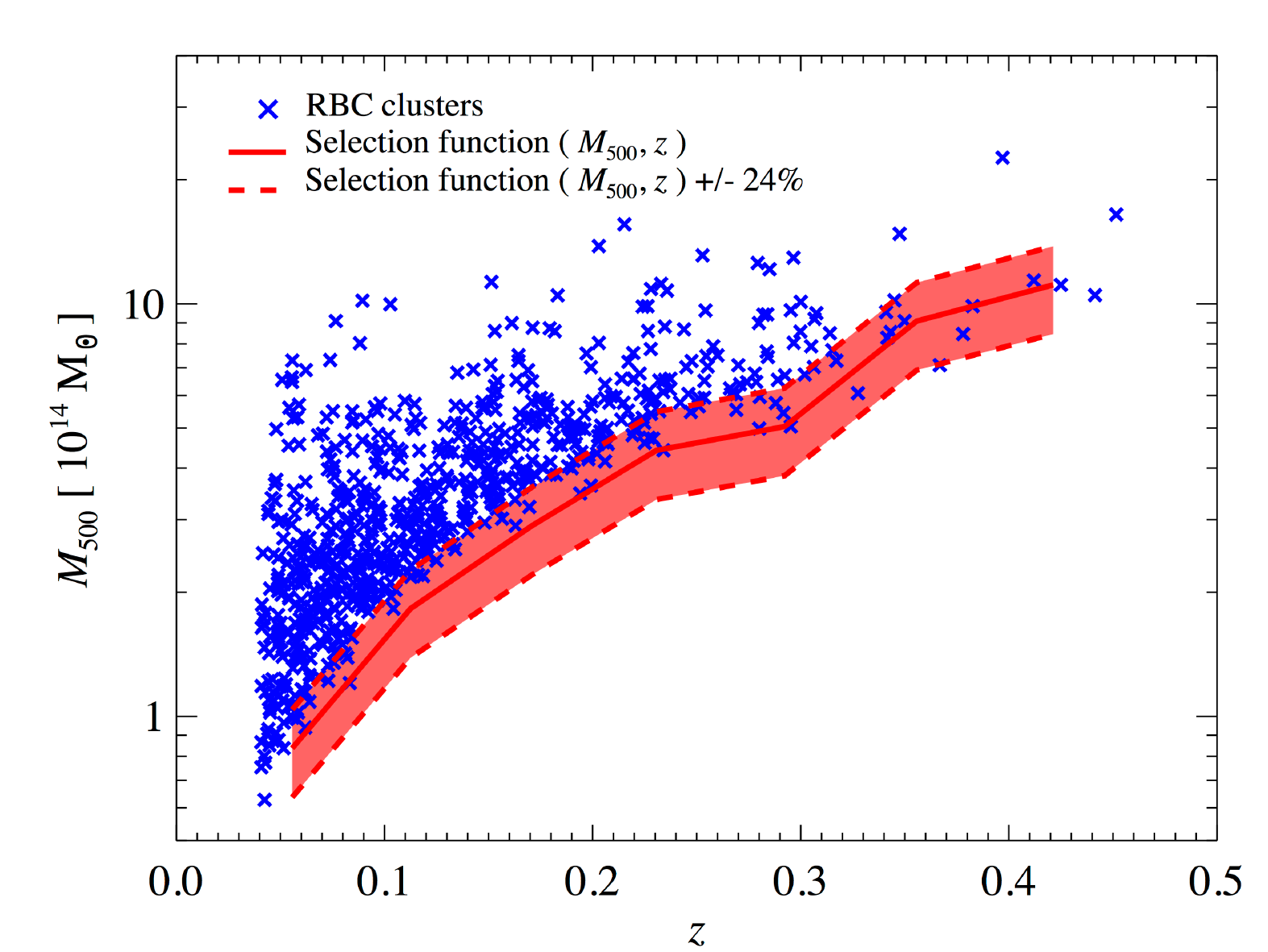}
\caption{The mass and redshift distribution of the RBC cluster catalog (Blue crosses). Plotted on top of the clusters is the selection function and $\pm 1 \sigma$ scatter in the selection selection (solid and dotted red lines, respectively) that is used in this analysis.}
\label{fig:cat}
\end{figure}

In Figure \ref{fig:cat} we show our RBC catalog as a function of mass and redshift. The selection function, $\Theta(m,z)$, is shown by the solid line, which we fit to a smoothed function for theoretical calculations. We illustrate the effect of $\pm$24 \% scatter in the $L_\rmn{x}-M$ relation by scaling the $\Theta(m,z)$ up and down by this percentage.

\section{Method} \label{sec:meth}

There are various components in the microwave sky, each with its own frequency dependence: 
\be \label{eqn:cmbMap}
\Delta T(\hat{\mathbf{\theta}}) = T_\rmn{SZ} + T_\rmn{CMB}+ T_\rmn{CIB} + T_\rmn{fg} + T_\rmn{PS} + N, 
\ee
where $T_{SZ}$,  $T_{CMB}$ and  $T_{CIB}$ represent the signal for the SZ clusters, the CMB and the cosmic infrared background (CIB). $T_{fg}$ represents all foregrounds including galactic dust, free-free and synchrotron, $T_{PS}$ shows radio galaxies and $N$ is the pixel noise which dominates at small scales.

Using the X-ray cluster catalog we generate two maps: a number-count map and a normalized overdensity map. We make the number-count map ($n(\hat{\mathbf{\theta}})$) first by placing ones in the central pixel the locations of clusters and smooth with the Planck beam (see below for more details). The normalized overdensity map is derived by dividing the number-count map by its mean value ($\bar{n}$) and subtracting one:

\be 
\delta_{n}(\hat{\mathbf{\theta}}) = \frac{n(\hat{\mathbf{\theta}})}{\bar{n}} - 1. 
\ee
\noindent This, in general, can be shown to obey 
\be \label{eqn:overdensityMap}
\delta_{n}(\hat{\mathbf{\theta}}) = \sum_{i=1}^{N} a_i f(\hat{\mathbf{\theta}}-\hat{\mathbf{\theta}}_i) + \rm{Const.},
\ee
where $f(\hat{\mathbf{\theta}}-\hat{\mathbf{\theta}}_i)$ is the shape we attribute to the clusters in our overdensity map. In the absence of any smoothing (no beam, infinitely small pixels in the map), $f(\hat{\mathbf{\theta}}-\hat{\mathbf{\theta}}_i)$ is a Kronecker delta function. In our analysis we perform a beam smoothing in map space using Planck's beam convolution tool, \textit{EffConv}\footnote{\url{http://irsa.ipac.caltech.edu/data/Planck/release_1/software/}} \citep{effconv}, to smooth the map with the 143 GHz beam of Planck. We have tested other choices, e.g., no beam smoothing and a top-hat pixel window function smoothing alone, and the final result is robust against the choice of the function $f$. There are various ways of choosing the weights $a_i$. In this analysis we choose uniform constant weights $a_i = a$.  

We measure the cross-power spectra of the maps in eqn. \ref{eqn:cmbMap} and both eqn. \ref{eqn:overdensityMap} and $n(\hat{\mathbf{\theta}})$. The difference between the cross-spectra will be an overall normalization factor proportional to $\bar{n}$. As we will see later on, the different maps ($\delta_{n}$ and $n$) are particularly sensitive to both cluster physics and the cosmological parameters governing the growth of structure such as $\sigma_8$ and $\Omega_\rmn{M}$.

Hereafter, we will mostly refer to the overdensity map but the same statements apply to the number-count map. Several different terms contribute to the cross-spectrum. The overdensity map made from the number-count map of the X-ray selected clusters is not correlated with the foregrounds or the CMB or noise in the Planck and WMAP sky maps. While, there is a small correlation between the overdensity map and the lensed CMB map, it is small enough to be ignored in this analysis. Although both clusters and the CIB trace the underlying mass distribution \citep[e.g.,][]{cosmicWeb, IxC}, the cluster catalog used in our analysis is a low-redshift catalog whereas the CIB  predominantly comes from high-redshift dusty star forming galaxies \citep[e.g.,][]{Marco}. Therefore, we do not expect a significant cross-correlation between the overdensity map and the CIB at the frequencies studied in this analysis. We explicitly check this by cross-correlating the 217 GHz map, the map with the strongest CIB signal, with the overdensity map.  Since this term is small, the SZ map-overdensity term dominates the cross-correlation: 

\be
C_{\ell}^{\Delta T \times \delta_{n}} \simeq C_{\ell}^{SZ \times \delta_{n}}.
\ee

\noindent The same applies for cross-correlation between number-count map and the CMB map:

\be
C_{\ell}^{\Delta T \times n} \simeq C_{\ell}^{SZ \times n}.
\ee

To compute the cross-power spectra, we use PolSpice\footnote{\url{http://www2.iap.fr/users/hivon/software/PolSpice/}} which deals with the mask effects in the map-space where the mode-mode correlation matrix is diagonal \citep{polspice}. Although these cross spectra are very similar their measurements constrain  different cluster quantities. The $C_{\ell}^{\Delta T \times \delta_{n}}$ spectrum measures the average SZ signal from the cluster sample, which constrains the average ICM properties of these cluster. The $C_{\ell}^{\Delta T \times n}$ spectra measure the total SZ signal from the cluster sample. This signal is not only dependent on the ICM properties but is very sensitive the number counts of clusters for a given cosmology. Thus, the $C_{\ell}^{\Delta T \times n}$ spectrum is a growth of structure measurement and can be used to constrain cosmology.

\section{Theoretical Power Spectrum Calculations} \label{sec:thry}

The theoretical predictions for the cross-correlation signal $C_{\ell}^{SZ \times \delta_{n}}$ and $C_{\ell}^{SZ \times n}$ were made using both an analytic halo model and cosmological simulations. We used the simulation approach to calculate the overall amplitude and shape of both $C_{\ell}^{SZ \times \delta_{n}}$ and $C_{\ell}^{SZ \times n}$. In the simulation predictions we included the scatter associated with the selection function using Monte Carlo method \citep{MC1949}. The analytical halo model was used to determine the cosmology dependence of $C_{\ell}^{SZ \times n}$. We used a similar halo model to ref. \citep{Dieg2003} with an updated cluster pressure profile \citep{Batt2012b} and included the selection function and the measured scatter for the RBC X-ray cluster catalog.

\subsection{Halo Model}

We constructed a halo calculation for the expected signal from this cross correlation $C_{\ell}^{SZ \times n}$ following ref. \citep{KS2002} and \citep{Dieg2003}. There are two components that contribute to $C_{\ell}^{SZ \times n}$, which can be expressed as a sum

\be
C_{\ell}^{SZ \times n} = C_{\ell,1\rmn{h}}^{SZ \times n} + C_{\ell,2\rmn{h}}^{SZ \times n},
\ee

\noindent where $C_{\ell,1\rmn{h}}^{SZ \times n}$ and $C_{\ell,2\rmn{h}}^{SZ \times n}$ are the 1-halo and 2-halo contributions, respectively. Similar to the tSZ auto-power spectrum \citep{KK1999}, the 1-halo term dominates $C_{\ell}^{SZ \times n}$ on the angular scales that we are modeling \citep[$\ell > 100$;][]{Dieg2003}. For this analytic calculation we ignore the contribution from the 2-halo term. For $C_{\ell}^{SZ \times \delta_{n}}$ we need to consider the amplitude and shape of tSZ signal. The X-ray clusters are modeled as delta functions convolved with the beam, thus, they do not require shape or amplitude modeling. The strength of the tSZ signal is proportional to the integrated electron pressure along the line of sight ($l$),

\be \label{eq:y}
y = \frac{\sigma_\rmn{T}}{m_\rmn{e}c^2}\int n_\rmn{e}kT_\rmn{e}\dd l,
\ee

\noindent where $\sigma_\rmn{T}$ is the Thompson cross section, $m_\rmn{e}$ is the electron mass, $c$ is the speed of light, $k$ is Boltzmann's constant, $n_\rmn{e}$ is the free electron density and $T_\rmn{e}$ is the electron temperature. The tSZ signal is $\Delta T / T_\rmn{CMB} = f_{\nu} y$, where $f_{\nu}$ is the spectral function for the tSZ signal \citep{Sunyaev:1972}.

The complete halo calculation for cross-power spectrum is given by 

\be \label{eq:theory}
C_{\ell}^{SZ \times n} = f_{\nu} \int_{0.04}^\infty \frac{\dd V}{\dd z} \dd z  \int_0^\infty \frac{\dd n(M,z)}{\dd M} \tilde{y}_{\ell}(M,z) \Theta(M,z) \dd M,
\ee

\noindent where $\dd n(M,z) / \dd M$ is the mass function, $\tilde{y}_{\ell}(M,z)$ is the Fourier transform of the projected electron pressure
profile (Eq. \ref{eq:y}) and $\Theta(M,z)$ is the selection function of the RBC catalog, which is described in Section \ref{sec:clsl}.  We do not include higher order relativistic corrections to $f_{\nu}$ \citep{Noza2006}. The mass function is determined from N-body simulations \citep[e.g.,][]{2001MNRAS.321..372J,2006ApJ...646..881W,2008ApJ...688..709T}. In this work we use the mass function from ref. \citep{2008ApJ...688..709T} for the analytic calculations and convert the definition of virial mass used in the RBC to the definition in ref.  \citep{2008ApJ...688..709T} which is defined with respect to the mean matter density. Because the tSZ power spectrum is  dominated by $10^{13}$- $10^{15}$ solar mass scale, it is only mildly sensitive to the particulars of the mass function \citep{KS2002}.

Using this analytic model, we calculate how the cross spectrum depends on two key concordance $\Lambda$CDM cosmological parameters, $\sigma_8$ and $\Omega_\rmn{M}$. 
These two parameters affect the number of clusters and their redshift dependence, a connection that has been explored for over two decades \citep[see e.g.,][]{Fan}. 
We choose the WMAP9 cosmological parameters as the baseline parameters and determine the scaling relationship.  The baseline cross spectrum is $C_{\ell}^{SZ \times n}(\sigma_8=0.8)$ and we compute new spectra by varying $\sigma_8$, $C_{\ell}^{SZ \times n}(\sigma_8)$, while holding the other parameters fixed. At a given angular scale we compare the relative amplitudes ($A_{\times}(\ell)$) of the spectra,

\be \label{eq:asig}
A_{\times}(\ell) = \frac{C_{\ell}^{SZ \times n}(\sigma_8)}{C_{\ell}^{SZ \times n}(\sigma_8=0.8)} \equiv \left(\frac{\sigma_8}{0.8}\right)^{\alpha_{\ell}}
\ee

\noindent which we assume scales as a power-law function of $\sigma_8$, where $\alpha_{\ell}$ is the power-law index for a given angular scale. The value of the power-law index is,

\be \label{eq:al}
\alpha_{\ell} = \frac{\dd\,\rmn{ln}A_{\times}(\ell)}{\dd\,\rmn{ln}(\sigma_8/0.8)}.
\ee

\begin{figure*}
  \resizebox{0.5\hsize}{!}{\includegraphics{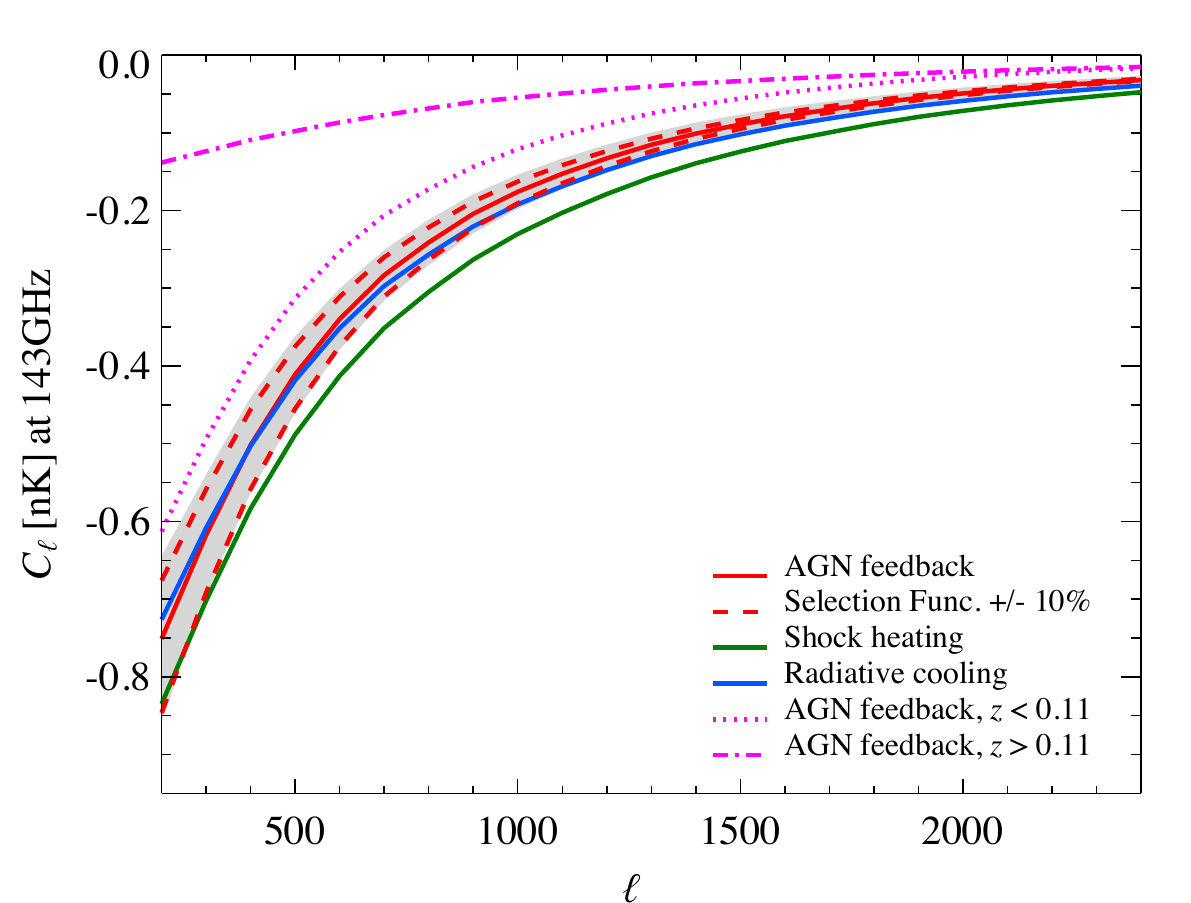}}%
  \resizebox{0.5\hsize}{!}{\includegraphics{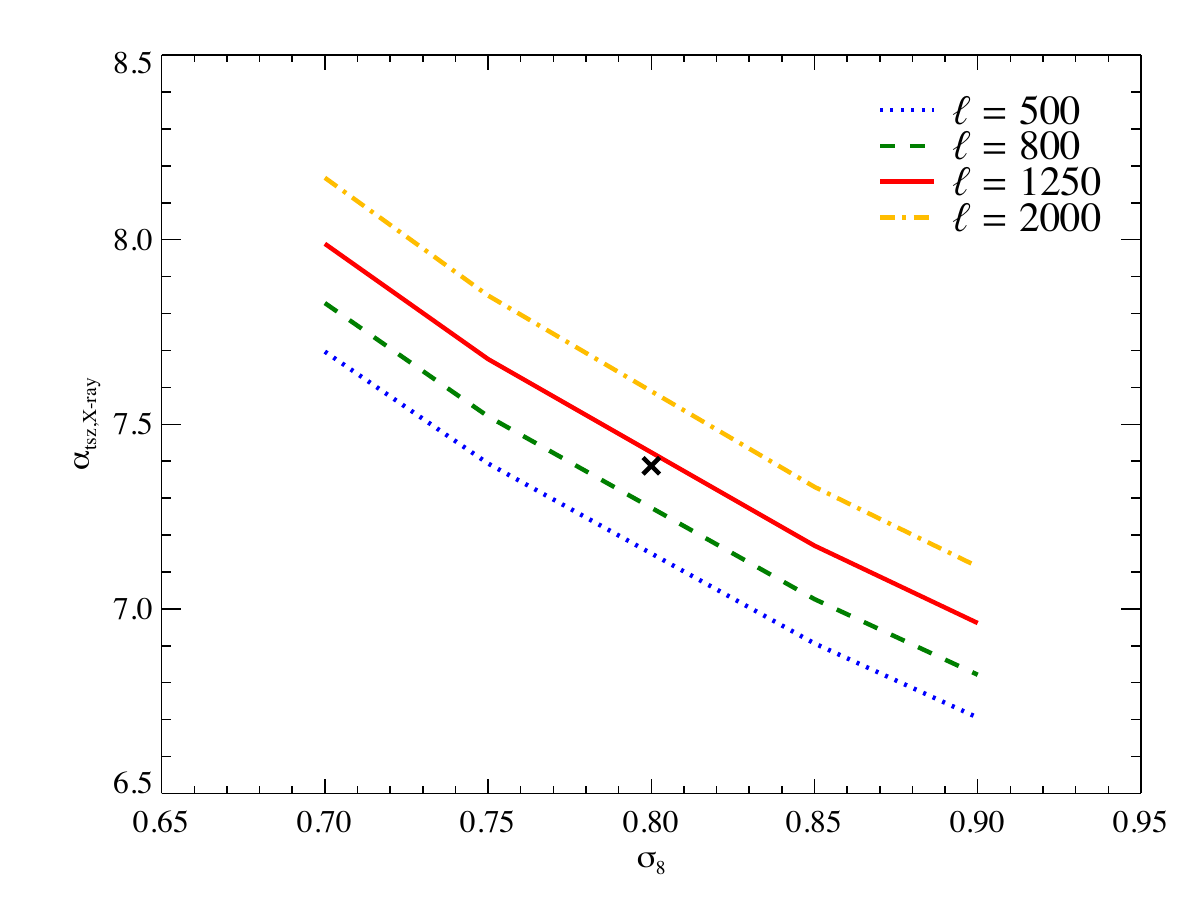}}\\ 
  \caption{\textit{Left:} The modeling uncertainties in the cross spectrum $C_{\ell}^{SZ \times \delta_{n}}$ from the ICM modeling ({\it gastrophyiscs}) and the selection function. We show the mean cross spectra for the three simulated ICM models {\it shock heating} (green line), {\it AGN feedback} (red line),  {\it radiative cooling} (blue line) and two variations of the selection function $\pm$ 10\% in mass (dashed lines) on the {\it AGN feedback} model. Additionally, for the {\it AGN feedback} we show the sample variance (grey band) and the contribution to the cross spectra from $z < 0.11$ (pink dotted line) and $z > 0.11$ (pink dot-dashed line). \textit{Right:} The power-law scaling of the cross spectrum with $\sigma_8$ as a function of $\sigma_8$ for $\ell = 500, 800, 1250, 2000$. The cross represents the $\alpha_{\ell} = 7.4$, which we use in our analysis. }
\label{fig:gastro}
\end{figure*}

\noindent In Figure \ref{fig:gastro} we show $\alpha_{\ell}$ for a range of $\sigma_8$ values from 0.7 to 0.9 at the angular scales $\ell = 500, 800, 1250$ and 2000. There is little $\ell$ dependence and $\alpha_{\ell} \sim 7.8$ to 6.9 spans the current measurements of $\sigma_8 = 0.82\pm 0.03$ \citep[e.g.,][]{Planck16} from CMB measurements. We set $\alpha_{\ell} = 7.4$ for the rest of this work.

The cross spectrum can be thought of as a resolved tSZ auto spectrum, which helps in understanding why $A_{\times}$ is highly dependent on $\sigma_8$. It is well established that the tSZ auto spectrum is a strong function of $\sigma_8$ \citep[e.g.,][]{KS2002,Shaw2010,Trac2011} and the cross spectrum is similar to tSZ auto spectrum except for two key differences: First, the cross spectrum is weighted\footnote{these weightings hold for objects that are unresolved at the angular resolution of interest, which is the case for most of the clusters in the BCS catalog} by $\tilde{y}_{\ell}(M,z) \propto M^{5/3}$ instead of $|\tilde{y}_{\ell}(M,z)|^2 \propto M^{10/3}$ which makes the amplitude of the auto spectrum a stronger function of $\sigma_8$ than the cross spectrum. Second, the selection function in the cross spectrum restricts the mass integration to rare and massive objects that populate the exponential tail of the mass function instead of the entire halo population. These two competing effects almost cancel each other yielding a $\sigma_8$ scaling for the cross spectrum that is similar to the auto spectrum.

A similar power-law index was calculated for $\Omega_\rmn{M}$\footnote{We assume that $\Omega = 1$, therefore any scaling with $\Omega_\Lambda$ will be the same as $\Omega_\rmn{M}$} using Eq. \ref{eq:al} replacing $\sigma_8$ with $\Omega_\rmn{M}$. We found an index for $\Omega_\rmn{M}$ of $1.9$, which is smaller than the $\sigma_8$ scaling. As was found with analyses of the mass function \citep[e.g.,][]{BBKS1986,Vik2009}, the  $C_{\ell}^{SZ \times n}$ signal depends weakly on the other cosmological parameters over their current uncertainty ranges.
For our cosmological analysis we only consider the dependence on $\sigma_8$ and $\Omega_\rmn{M}$ when converting the amplitude of $C_{\ell}^{SZ \times n}$ to cosmological constraints. The complete relationship between $\sigma_8$, $\Omega_\rmn{M}$ and $A_{\times}$ is

\be \label{eq:asig_all}
A_{\times} =  \left(\frac{\sigma_8}{0.8}\right)^{7.4} \left(\frac{\Omega_\rmn{M}}{0.30}\right)^{1.9}.
\ee

\noindent We account for the small $\ell$ dependence of $A_{\times}$ by averaging over the $A_{\times}$ values bewteen $\ell = 900 - 1300$, where we have the largest signal to noise across the 100, 143 and 353 GHz spectra (see Sec. \ref{sec:concl}). It is obvious from Eq. \ref{eq:asig_all} that in this analysis any measurement of $A_{\times}$ will yield degenerate constraints on $\sigma_8$ and $\Omega_\rmn{M}$. In order to break this degeneracy we use external primary CMB parameter chains from WMAP and Planck to constrain the values of $\sigma_8$ and $\Omega_\rmn{M}$.

Additionally, we calculated the theoretical variances using our halo model by computing the diagonal of the covariance matrix

\be \label{eq:covar}
M_{\ell\ell}^{SZ \times n} = \frac{1}{f_\rmn{sky}}\left[\frac{2 (C_{\ell}^{SZ \times n})^2}{2\ell+1} +\frac{T_{\ell\ell}^{SZ \times n}}{4\pi}\right], 
\ee

\noindent where $f_\rmn{sky}$ is the fraction of the unmasked sky and $T_{\ell\ell}^{SZ \times n}$ is the angular trispectrum for the cross correlation \citep[for more details on analytical tSZ auto spectrum errors see][and references therein]{KS2002}. The trispectum represents the {\it non-Gaussian} error term and is defined as 

\be \label{eq:trispec}
T_{\ell\ell}^{SZ \times n} = f_{\nu}^2 \int_{0.04}^\infty \frac{\dd V}{\dd z} \dd z  \int_0^\infty \frac{\dd n(M,z)}{\dd M} |\tilde{y}_{\ell}(M,z)|^2 \Theta(M,z) \dd M .
\ee

\noindent Similar to the tSZ autospectrum the diagonal of Eq. \ref{eq:covar} is dominated by the $(C_{\ell}^{SZ \times n})^2$ term (i.e. the {\it Gaussian} error term) and not the trispectrum. Thus, the analytical fractional errors for $C_{\ell}^{SZ \times n}$ ($\Delta C_{\ell}^{SZ \times n}/C_{\ell}^{SZ \times n} \equiv \sqrt{M_{\ell\ell}^{SZ \times n}}/C_{\ell}^{SZ \times n}$) scale like $1/\sqrt{\ell}$ on the angular scales we are interested in. As we discuss in section \ref{sec:ThryUn}, these errors are smaller than what we find using the simulations.

\subsection{Numerical Predictions} \label{sec:num}
Using a modified version of the GADGET-2  smoothed particle hydrodynamics (SPH) code \citep{Gadget}, we simulate large cosmological volumes ($L=165$ Mpc$/h$) so that we have a fair statistical sample of halos for  calculating the cross spectrum. 
The modified version of the GADGET-2 code \citep{Gadget} we used included additional sub-grid physics models for active galactic nuclei (AGN) feedback \citep[for more details see][]{Batt2010}, radiative cooling and star formation \citep{SpHr2003}. We use three variants of sub-grid physics models in the simulations: the first is a non-radiative case with only gravitational heating ({\it shock heating}); the second is a model that accounts for radiative cooling, star formation, supernova feedback, cosmic ray physics \citep[for more details see][]{2006MNRAS.367..113P,2007A&A...473...41E,2008A&A...481...33J} and galactic winds ({\it radiative cooling}); the third is the full {\it radiative cooling} model with an additional model for AGN feedback ({\it AGN feedback}). The {\it shock heating} model shows significant discrepancies with cluster observations \citep[e.g.,][]{Puch2008}. We do not present them as a viable alternative to the {\it AGN feedback} model but as an extreme example of an ICM model. There are a total of 10 simulations for each physics model. Each simulation has a box size of 165 Mpc$/h$, 256$^3$ gas and dark matter (DM) particles with a mass resolution of $M_\rmn{gas} = 3.2\times 10^{9} \rmn{M}_{\odot}/h$ and $M_\rmn{DM} = 1.54\times 10^{10} \rmn{M}_{\odot} /h$. The cosmology used in the simulations was $\Omega_\rmn{M} = \Omega_\rmn{DM} + \Omega_\rmn{b} = 0.25$, $\Omega_\rmn{b} = 0.043$, $\Omega_\Lambda = 0.75$, $H_0=100\,h\,\rmn{km}\,\rmn{s}^{-1}\,\rmn{Mpc}^{-1}$, $h=0.72$, $n_\rmn{s} =0.96$ and $\sigma_8 = 0.8$. 

We identify halos and calculate their properties in two steps. First the halos are found using a friends of friends algorithm \citep{Huch1982}. Second, for each halo we compute the center of mass iteratively and then the spherical overdensity mass ($M_{\Delta}$) and radius ($R_{\Delta}$). Here $\Delta$ refers to the multiplicative factor applied to the critical density, $\rho_\rmn{cr}(z) \equiv 3 H_0^2 [\Omega_\rmn{M}(1+z)^3 + \Omega_{\Lambda}] / (8\pi\,G)$. 

At each simulation output that corresponds to a redshift, we make three maps. The first map is a number-count map with  a non-zero value at the location of each halo above a given redshift dependent minimum mass.  As part of the cluster selection,  we scale each cluster mass by a randomly selected number that is drawn from a Gaussian distribution with a mean of 1 and a dispersion that matches the $L_\rmn{X}-M$  scatter of 24\% and then determine whether the``observed" mass, the product of this random number times the true mass, is above the redshift dependent cutoff. 
We make 200 random realizations of a number-count map for each simulation, totaling 2000 maps per redshift. The second map is an overdensity map ($\delta$). These maps are the number-count maps that have been normalized by the ratio of the total number of halos in the lowest mass bin over the number of pixels in the map.
The third map is the Compton-y map, described in Eq. \ref{eq:y}. We cross correlate the number-count and overdensity maps with the Compton-y  at each simulation output for all the simulations above various cluster masses. Then for each redshift and mass bin we average over all simulations realization and calculate the variance. The total signal is calculated by summing up the appropriate mass bin at the given redshift according to the selection function. Both $C_{\ell}^{SZ \times n}$ and $C_{\ell}^{SZ \times \delta_{n}}$ have the same shape and only differ in their amplitudes.

\subsection{Theoretical Uncertainties} \label{sec:ThryUn}

The uncertainty on the theoretical amplitude of $C_{\ell}^{SZ \times n}$ can be separated into two categories, ICM properties ({\it gastrophysics}) and selection function uncertainties. The {\it gastrophysical} uncertainties of the ICM are found in the $\tilde{y}_{\ell}(M,z)$ of Eq. \ref{eq:theory}. There is only one power of $\tilde{y}_{\ell}(M,z)$, unlike the tSZ auto power spectrum which is $\propto |\tilde{y}|^2$ (i.e., they have different form factors). Thus, the cross power spectrum is less sensitive to {\it gastrophysics} than the auto spectrum. The uncertainties on $\Theta(M,z)$ impact the limits of the integration for $C_{\ell}^{SZ \times n}$. These limits are set by the calibration of the $L_\rmn{x}-M$ relation. Any biases in the $L_{\rmn{x}}-M$ relation will effect the amplitude and shape of the cross power. We explore both these uncertainties with simulations. 

Figure \ref{fig:gastro} shows the dependence on the different physics models of the simulations and variations in the selection function. The {\it shock heating} model has a larger signal than the {\it AGN feedback} model since the ICM in the {\it shock heating} does not lose baryons to star formation. The {\it radiative cooling} model has a similar signal as the {\it AGN feedback}. At low $\ell$, where the clusters are unresolved, the {\it radiative cooling} model has less power than the {\it AGN feedback} since the ICM over-cools, forms too many stars (i.e., removes too many baryons) and lowers the total thermal energy in the clusters. At higher $\ell$, the clusters are resolved and the {\it radiative cooling} model has more power than {\it AGN feedback} since the clusters in this model have higher thermal pressure in their interiors. This result is also found in the tSZ auto power spectrum \citep{Batt2010}. These three models have similar shapes, so departures from a more realistic ICM physics model ({\it AGN feedback}) to the most extreme model ({\it shock heating}) accounts for $\sim 30$\% difference in amplitude ({\it radiative cooling} $\sim 10$\% difference) over the $\ell$ range of interest. A measurement of $C_{\ell}^{SZ \times \delta_{n}}$ will help exclude such extreme ICM models (see Sec. \ref{sec:results}). Additionally, we split the cross spectrum into two redshift bins $z > 0.11$ and $z < 0.11$. For $\ell < 2500$ the cross spectrum is dominated by low-$z$ clusters (see Fig. \ref{fig:gastro}), which is purely a resolution effect. The central pixels of low-$z$ clusters are brighter than high-$z$ clusters. For $\ell > 2500$ the high-$z$ clusters dominate the cross spectrum signal.

The scatter in the $L_\rmn{X}-M$ is already accounted for using a Monte Carlo method (see Sec \ref{sec:num}). The other major uncertainty in the analysis 
is the overall bias in the $L_\rmn{X}-M$ relation.
For our baseline model, we increase the  hydrostatic mass by  20\%.  However, there is a 40\% uncertainty in the amplitude of this correction \citep[see][for a review]{Rasia2012}.  To model this uncertainty, we apply mass corrections at both extremes (10\% and 30\% ), propagate it through our simulation of the  selection function and recalculate the cross spectrum on the {\it AGN feedback}  simulations. We refer to these two cases as -10\% and +10\% throughout the rest of the paper.

In Figure \ref{fig:gastro} we show the impact of changing the selection function by this uncertainty. These extreme biases account for a $\sim \pm$ 7\% change in the cross spectrum amplitude and no change in the shape. Thus, it is not possible for the uncertainty in the $L_\rmn{x}-M$ mass calibration to compensate for the large differences in $C_{\ell}^{SZ \times \delta_{n}}$ amplitudes between ICM models.

In addition to the cross spectrum modeling uncertainties, we include the uncertainty from sample variance to the overall spectrum errors when we fit for the amplitude. We use the variances calculated from the simulations ($\sim 14$\% see Fig. \ref{fig:gastro}), since they are larger than the variances from the analytic error calculation ($\lesssim 10$\% see Eq. \ref{eq:covar}).

\section{Results}
\label{sec:results}
In our analysis we use one map from WMAP data at 96 GHz and six maps from Planck at 100, 143, 217, 353, 545 and 875 GHz as described in Section \ref{sec:data}. We compute the cross spectra between each of these maps and the RBC overdensity and number-count maps. Results for the $C_{\ell}^{SZ \times \delta_{n}}$ are presented in Fig. \ref{fig:wxr} and they show a significant correlation between the clusters and the microwave maps. The cross spectra qualitatively have the same shape as the theory curves shown in Fig. \ref{fig:gastro}. Moreover, the cross spectrum of the cluster overdensity map with WMAP 94 GHz (green data points) is consistent with the cross spectrum using the Planck 100 GHz map (red dots). The error bars in our analysis are computed analytically and we have verified them using Monte Carlo randomization of the phases. There are two sources of uncertainties in the measured cross spectra: the cosmic variance and the noise\footnote{There is a term in the analytic estimation of the errorbars that corresponds to the Poisson uncertainties due to finite cluster numbers. This component is dominant on small angular scales. On the scales of interest in our study the dominant error comes from the Gaussian uncertainties and on small scales pixel noise uncertainty dominates. The agreement between the analytical estimates using only the Gaussian part and the simulations suggests that the Poisson errors are small compared to other sources of uncertainty in the measured cross spectra. For a detailed description of the uncertainties on cross-spectra see the Appendix of \cite{AxB}. The theoretical uncertainties are discussed in Section \ref{sec:ThryUn}.}. The cosmic variance term is given by the power spectrum of the signal and is inversely proportional to the number of modes in each power spectrum bin. The noise term is given by the power spectrum of the noise (i.e., the CMB, foregrounds and the detector noise in Planck). The total uncertainty on the cross-spectrum is a superposition  of these two components. In addition, we include the sample variance from the theoretical model to the measured uncertainties.

\subsection{Constraints}
We fit a four-parameter model (an amplitude, scaled by the spectral function for the tSZ at each frequency, $f_\nu$, and Poisson terms for 100, 143 and 353 GHz) to the measured cross-spectra, $C_l^{\rmn{obs}}(\nu)$. The fit is parametrized using 
\be
C_l^{\rmn{obs}}(\nu) = A_{\times}(\nu) C_l(\nu_0) + B(\nu),
\ee

\noindent where the $C_l(\nu_0)$ is the baseline cross spectrum template calculated from simulations at the frequency $\nu_0$ and scaled to a given microwave frequency, $A_{\times}(\nu) = A_{\times} f_\nu/f_{\nu_0}$ is the normalization factor, and $B$ is the baseline resulting from the Poisson part of the power spectrum. In the absence of all uncertainties and systematics $A_{\times}$ should be identically 1. 

\begin{figure*}[t!]
\centering 
 \resizebox{0.5\hsize}{!}{\includegraphics{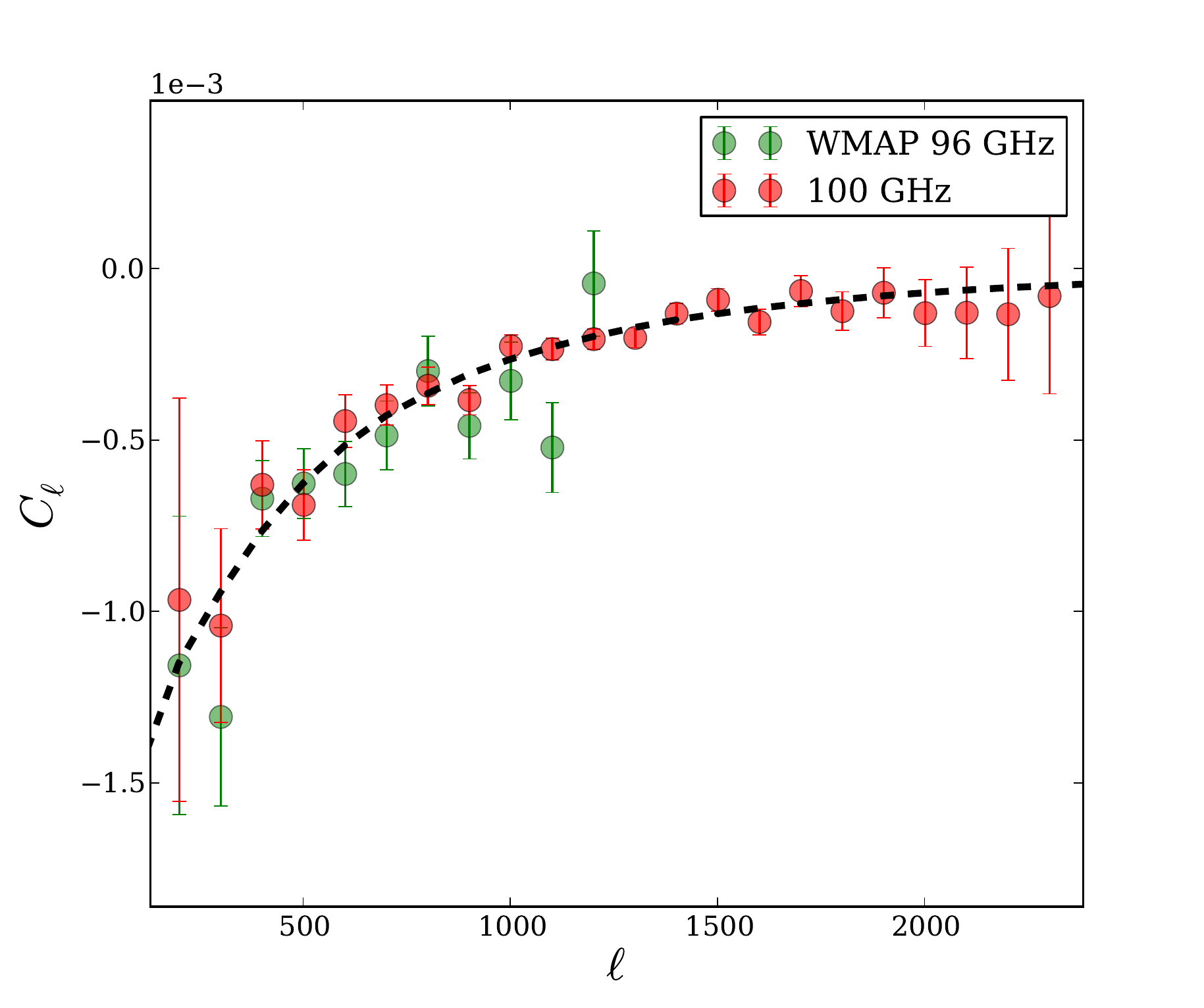}}%
 \resizebox{0.5\hsize}{!}{\includegraphics{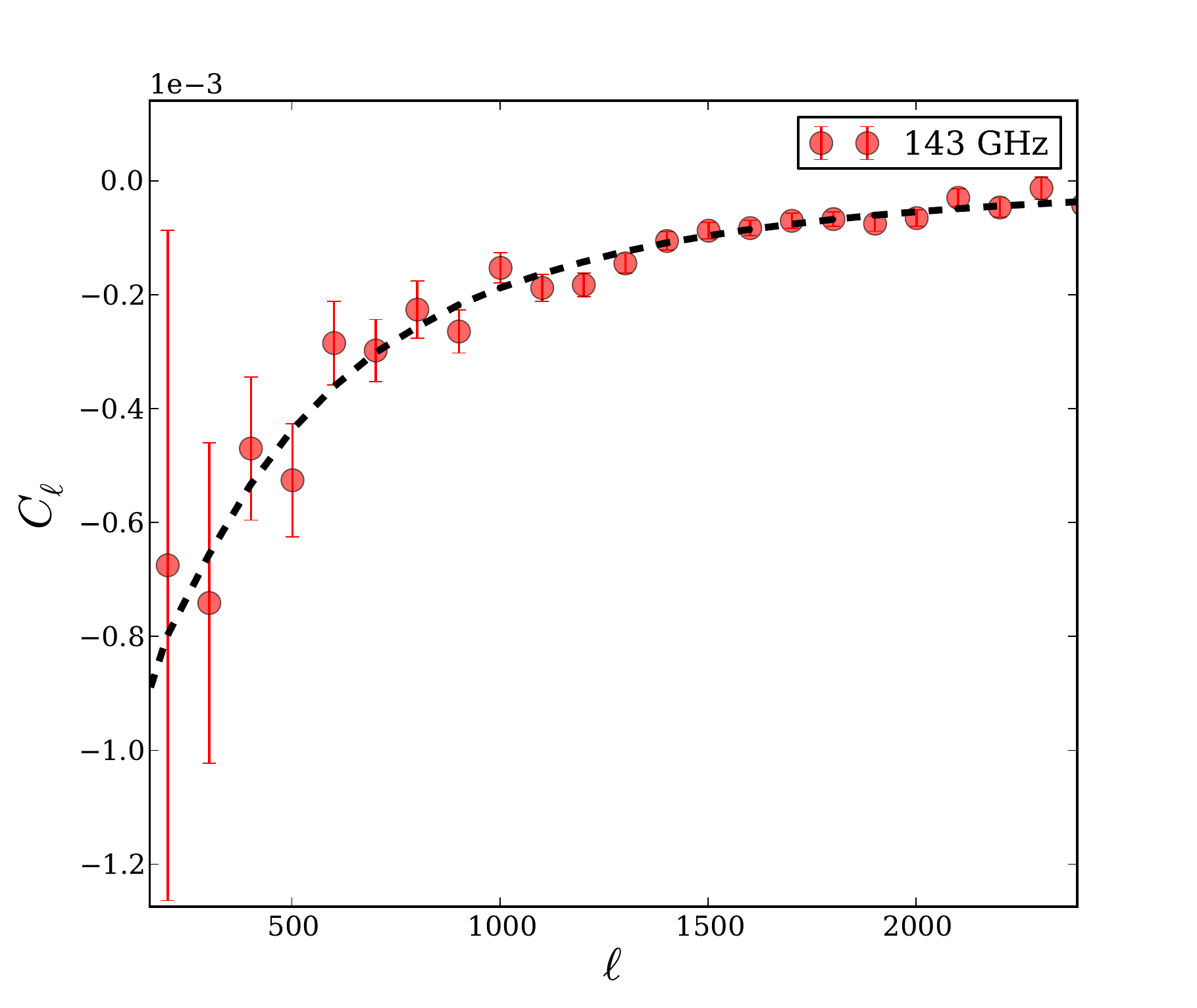}}\\
  \resizebox{0.5\hsize}{!}{\includegraphics{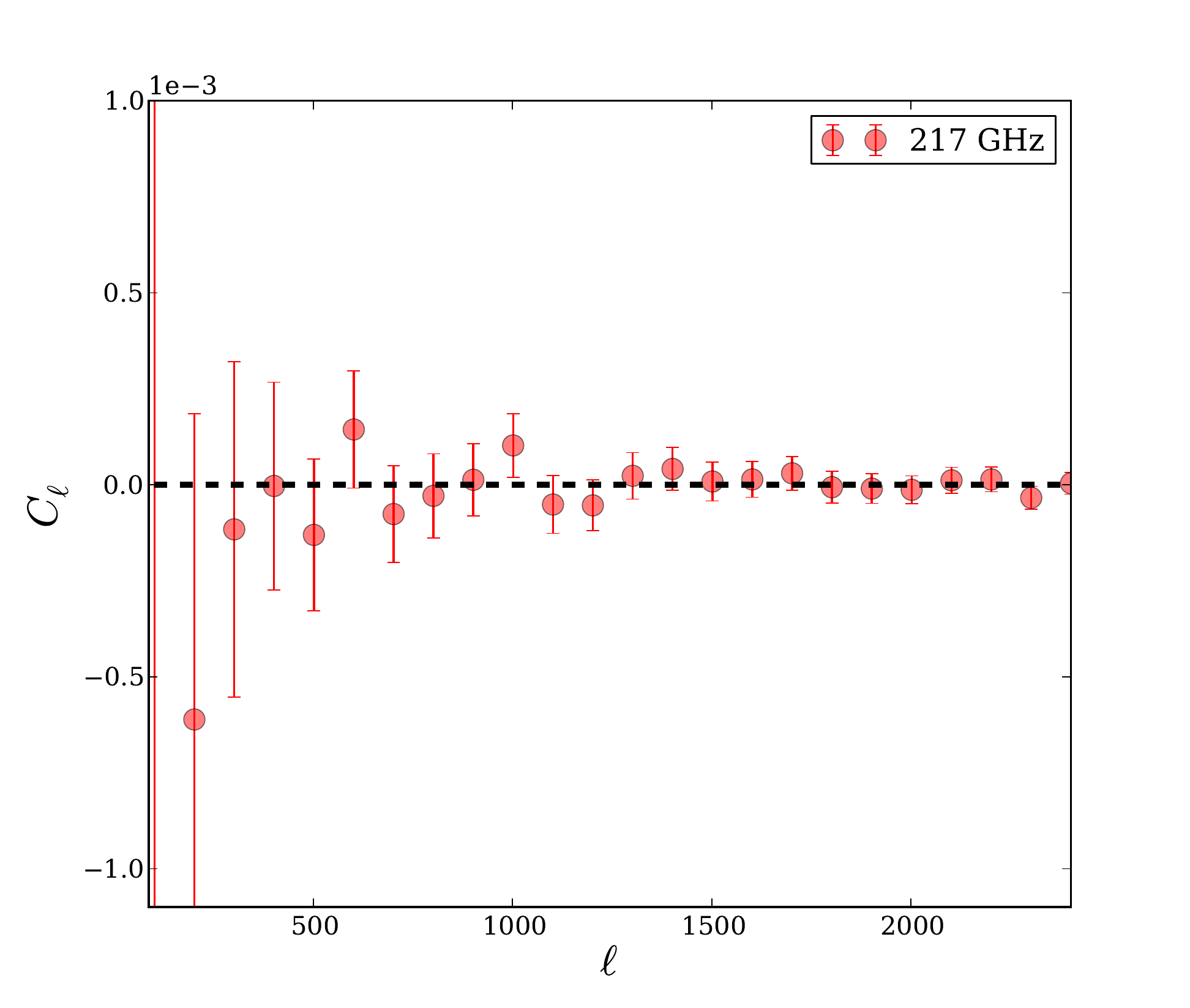}}%
 \resizebox{0.5\hsize}{!}{\includegraphics{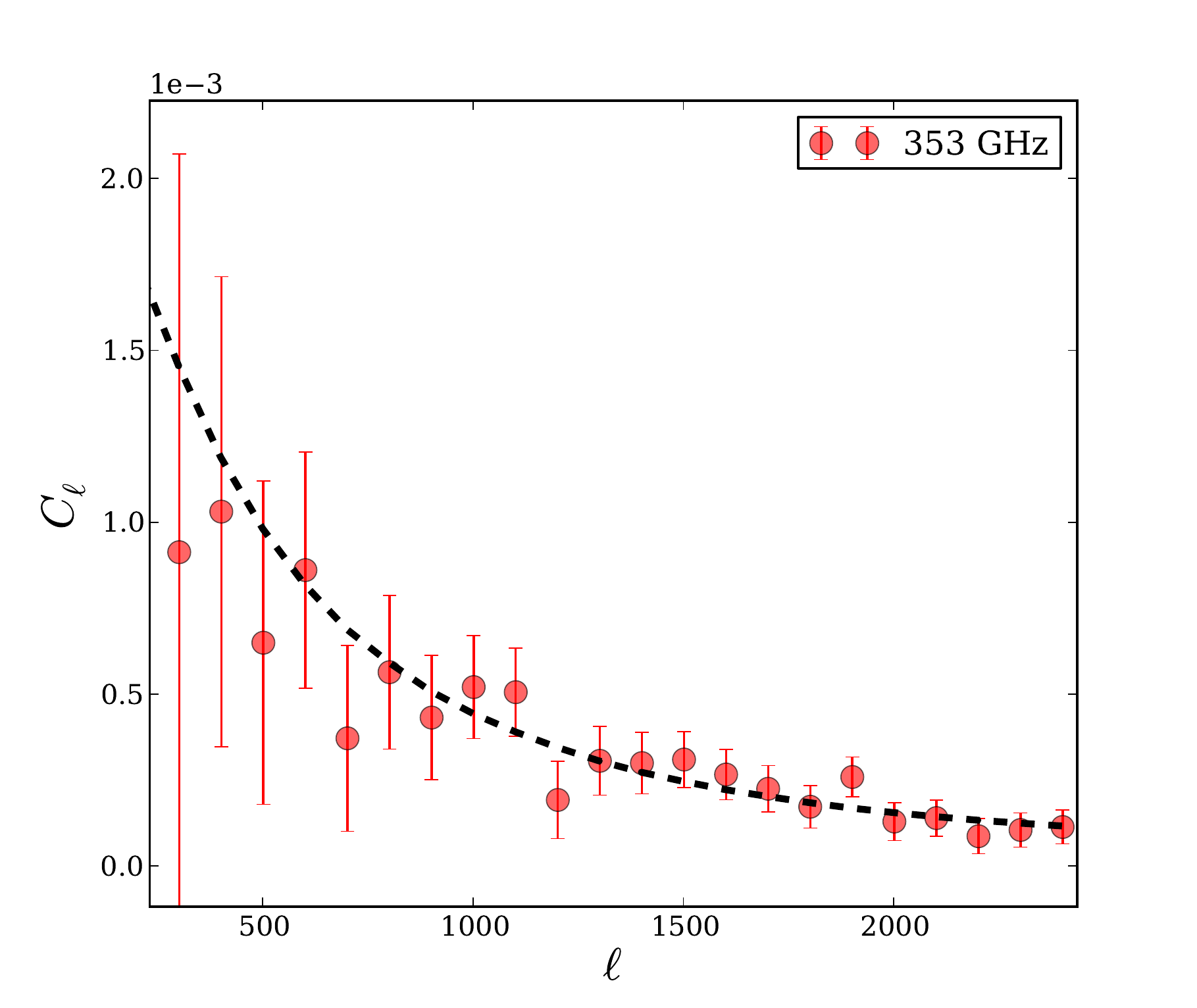}}\\
\caption{Cross-correlations of Planck and WMAP with ROSAT clusters. {\it Upper left:} Green data points show the overdensity cross-power spectrum of ROSAT and WMAP 96 GHz (W-band) in addition to Planck 100 GHz (red). The remaining panels show 143 GHz({\it upper right}), 220 GHz ({\it lower left}) and 353 GHz ({\it Lower right}) overdensity cross-spectra with ROSAT. The best fit theory model for all panels are shown by the black dashed line. The tSZ cluster signal is a decrement at frequencies below 220 GHz and an increment above that, hence the change in the sign of the cross-correlations at the highest frequency end. The number-count cross-spectra have exactly the same shape only the amplitudes differ, so they are not shown.}
\label{fig:wxr}
\end{figure*}

\begin{figure*}[t!]
\centering 
 \resizebox{0.9\hsize}{!}{\includegraphics{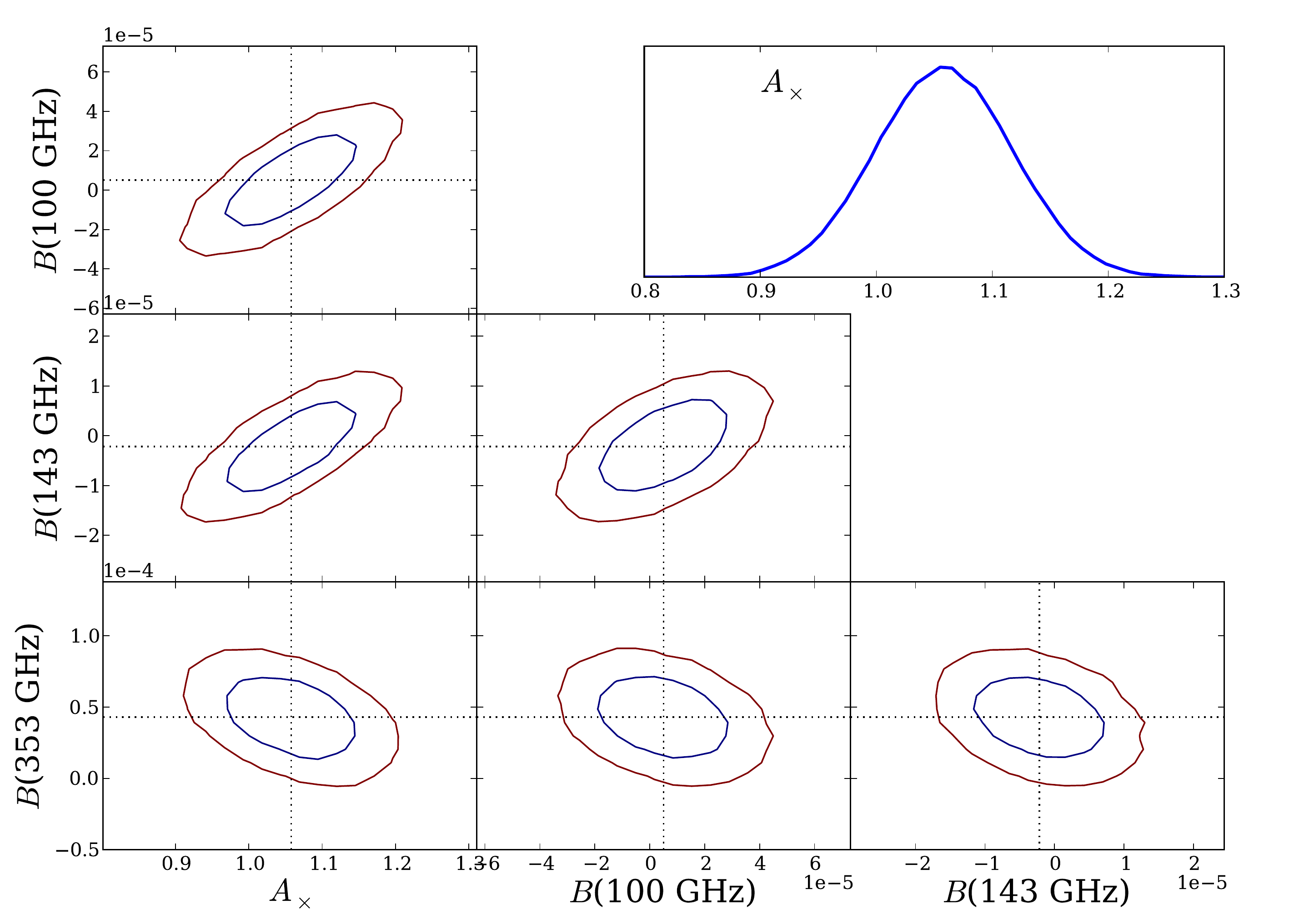}}%
\caption{Covariances of the model parameters, fit to the measured overdensity cross-spectra using MCMC. All the fit amplitudes can be found in Table \ref{tab:results}. \textit{Top right:} Shows the marginalized posterior of the normalization parameter, $A_{\times}$.}
\label{fig:params}
\end{figure*}

We use \textit{emcee}\footnote{\url{http://dan.iel.fm/emcee/}}, an efficient MCMC sampler publicly available in Python \citep{emcee}, to find the best fit model parameters. We fit all three frequencies (100, 143, 353 GHz) jointly. The results are shown in Fig. \ref{fig:wxr}. The dashed-line shows the best fit model and the red data points show the cross-spectra. We fit the amplitude of the measured cross spectra, both $C_{\ell}^{SZ \times \delta_{n}}$ and $C_{\ell}^{SZ \times n}$, against the templates from the {\it AGN feedback}, {\it shock heating} and {\it radiative cooling} models as well as the {\it AGN feedback} model with $\pm 10$\% uncertainty in the selection function mass calibration.

The best fit amplitude parameters ($A_{\times}$) for various models are given in Table \ref{tab:results}. We use the amplitude of $C_{\ell}^{SZ \times \delta_{n}}$ to constrain the ICM physics models since it is not sensitive to cosmology. Both the {\it AGN feedback} and {\it radiative cooling} models agree with the measured spectra and each other, while the {\it shock heating} model is over 2$\sigma$ discrepant with the measured spectrum. The uncertainty in the selection is not enough to reconcile the {\it shock heating} model with these measurements. The {\it AGN feedback} and {\it radiative cooling} models with a marginally higher mass biases match the overdensity cross spectra amplitudes the best. Thus, we find that the $L_\rmn{x}-M$ scaling relation (plus 20\% hydrostatic bias) is well represented by both the {\it AGN feedback} and {\it radiative cooling} models with minor corrections to the average hydrostatic bias value we assume. We find that increasing the mass bias by 10\% (+10\% model) results in a larger overdensity amplitude compared to the {\it AGN feedback model}, thus a lower fit value. This in the result of  the higher mass threshold reducing the amount of low mass clusters in the sample which increases the average tSZ signal. Decreasing the mass bias by 10\% (-10\% model) has the opposite effect. For the number-count spectra the +10\% model has a smaller amplitude and a larger fit value compared to the {\it AGN feedback model}. Here the removal of low mass clusters from the sample also removes signal.

For the cosmological constraints we use Eq. \ref{eq:asig} to convert the $A_{\times}$ fit values from the number-count spectra to $\sigma_8-\Omega_\rmn{M}$ values. Using the {\it AGN feedback} model we find $\sigma_8(\Omega_\rmn{M}/0.30)^{0.26} = 0.796 \pm 0.011$. The error bar on this constraint is determined entirely from the fit parameters without including the uncertainties on the selection function. After including primary CMB constraints from WMAP9 and Planck we obtain $\sigma_8$ constraints of $\sigma_8 = 0.812 \pm 0.009$ and $\sigma_8 = 0.810 \pm 0.007$ using the {\it AGN feedback} model, respectively. 

We obtain more realistic errors on $\sigma_8-\Omega_\rmn{M}$ by combining the posterior probabilities of the $A_{\times}$ fits that include the $\pm$ 10 \% uncertainty on the selection function. After combining these uncertainties the $\sigma_8-\Omega_\rmn{M}$ constraint becomes $\sigma_8  (\Omega_\rmn{M}/0.30)^{0.26} = 0.797 \pm 0.015$. We show the $\sigma_8$ and $\Omega_\rmn{M}$ contour that includes the uncertainty in the selection function in Figure \ref{ts}. After including primary CMB constraints from WMAP9 and Planck we obtain $\sigma_8$ constraints of $\sigma_8 = 0.812 \pm 0.010$ and $\sigma_8 = 0.812 \pm 0.008$ using the {\it AGN feedback} model plus selection uncertainty, respectively.

\subsection{Null and Cluster Catalog Tests}
The 217 GHz  map  is at the null-point of the tSZ signal and hence we use it for null tests. The bottom left panel of Figure \ref{fig:wxr} shows the cross-spectrum of this map with the cluster overdensity map. The reduced $\chi^2$ value for this cross-spectrum compared to zero is 0.91. In addition to being a null-test, this cross spectrum also constraints the dust contamination in the RBC clusters. Being at the null point of the tSZ signal, the main component in the 217 GHz map apart from the primordial CMB signal is the CIB component arising from the dusty star forming galaxies. The 217 GHz null cross-spectrum with the cluster overdensity map confirms that there is negligible dust contamination in our signal. Additionally, we use the two highest Planck frequency maps at 545 and 857 GHz to cross correlate with our cluster map. At these frequencies the dust signal dominates and the CMB signal is negligible. Figure \ref{fig:tests} shows the results. There is a large scatter in the cross-spectra with no significant evidence of a positive signal that hints towards correlations between the CIB at Planck frequencies and the RBC cluster catalog. These results are not surprising as the CIB in microwave frequencies is expected to primarily come from high-redshift dusty star forming galaxies \citep[DSFG's, ][]{AxB} whereas the RBC catalog is a low-redshift sample. Hence the absence of a significant correlation between the two is not unexpected. It also supports our assumption that a two component modeling of the cross-power spectra of Planck and the cluster overdensity map is sufficient.

\begin{table}[t!]\label{tab:results}\scriptsize 
\centering
\begin{tabular}{llcccc}
 \toprule[1.5pt]
\multicolumn{1}{c}{\head{Model}} & \multicolumn{1}{c}{\head{Spectrum}} &\multicolumn{1}{c}{\head{Fit Parameter}} & \multicolumn{3}{c}{\head{Derived Parameters}}\tabularnewline
\hline
& & $A_{\times}$ & $\sigma_8(\Omega_M / 0.3)^{0.26}$ & $\sigma_8$(WMAP9) & $\sigma_8$(Planck13) \tabularnewline
\cmidrule(lr){3-3}
\cmidrule(l){4-6}
\head{{\it AGN feedback}} & overdensity & 1.06 $\pm$ 0.06 & -- & -- & --\tabularnewline
& number-count  & 1.36 $\pm$ 0.14 &  0.796 $\pm$ 0.011 &  0.812 $\pm$ 0.009 & 0.810 $\pm$ 0.007 \tabularnewline
  \cmidrule(lr){1-6}
 \head{{\it radiative cooling}} & overdensity & 1.04 $\pm$ 0.06 &  -- & -- & --\tabularnewline
& number-count & 1.33 $\pm$ 0.14 & 0.793 $\pm$ 0.011 & 0.811 $\pm$ 0.009 & 0.809 $\pm$ 0.007 \tabularnewline
\cmidrule(lr){1-6}
\head{{\it shock heating}} & overdensity & 0.88 $\pm$ 0.05 &  -- & -- & --\tabularnewline
& number counts & 1.12 $\pm$ 0.12 & 0.775 $\pm$ 0.011 & 0.802 $\pm$ 0.009 & 0.801 $\pm$ 0.008 \tabularnewline
   \cmidrule(lr){1-6}
 \head{{\it AGN feedback} +10\%} & overdensity & 0.95 $\pm$ 0.05 &  -- & -- & --\tabularnewline
& number-count & 1.53 $\pm$ 0.15 & 0.809 $\pm$ 0.011 & 0.819 $\pm$ 0.009 & 0.815 $\pm$ 0.007 \tabularnewline
   \cmidrule(lr){1-6}
 \head{{\it AGN feedback} -10\%} & overdensity & 1.16 $\pm$ 0.07 &  -- & -- & --\tabularnewline
& number-count & 1.25 $\pm$ 0.13 & 0.787 $\pm$ 0.010 & 0.808 $\pm$ 0.009 & 0.805 $\pm$ 0.007 \tabularnewline
\bottomrule[1.5pt]
\head{{\it AGN feedback} all} & number-count & 1.38 $\pm$ 0.19 & 0.797 $\pm$ 0.015 & 0.812 $\pm$ 0.010 & 0.812 $\pm$ 0.008 \tabularnewline
\bottomrule[1.5pt]
\end{tabular}
\caption{Best fit values of the normalization parameter $A_{\times}$ and its fit uncertainty for the models and data sets used in this analysis. In the adjacent column we show the derived $\sigma_8(\Omega_M / 0.3)^{0.26}$ values from the fit parameter, $A_{\times}$, using Eq. \ref{eq:asig}. The errors on $\sigma_8(\Omega_M / 0.3)^{0.26}$ are computed using the fit uncertainties only, except for the bottom row where we included the $\pm 10$\% uncertainty on the selection function mass by combining the posterior probabilities. The actual error bars for the derived $\sigma_8$ values are asymmetric, but the asymmetry is small, so we quote symmetric error bars.
The $\sigma_8(\rmn{WMAP9})$ and $\sigma_8(\rmn{Planck13})$ columns are the $\sigma_8$ constraints after included the primary CMB constraints from WMAP and Planck (for data set details see Sec. \ref{sec:data})}
\end{table}

\begin{figure*}[t!]
\centering
\includegraphics[width=1.0\textwidth]{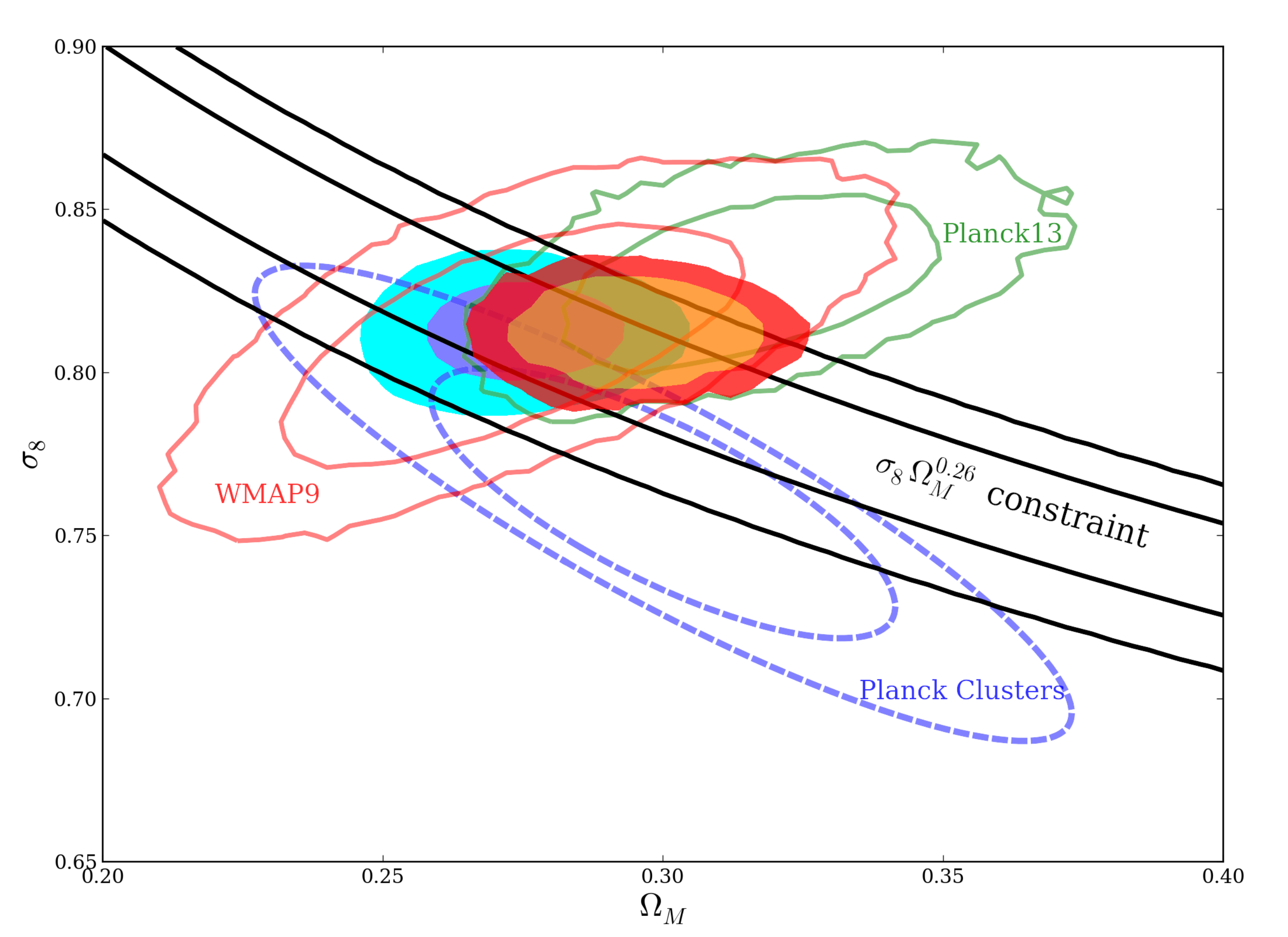}
\caption{Derived joint constraints on $\sigma_8$ and $\Omega_M$ parameters from the measurement of the cross-correlation normalization coefficient, $A_{\times}$. The black lines show 1- and 2-$\sigma$ levels for the AGN feedback model including all uncertainties. The red contours are from WMAP9 + SPT + ACT \citep{WMAP9cos} and the green contours show Planck + WMAP low $\ell$ polarization + high-$\ell$ \citep{Planck16}. The filled contours show the result of combining our measurements (the black band) with Planck (red filled contours) and with WMAP9 (blue filled contours). The dashed blue lines show the likelihood contours for Planck clusters + BAO + BBN from  \citep{Planck-XX}.}
\label{ts}
\end{figure*}

\begin{figure*}[t!]
\centering
\includegraphics[width=0.47\textwidth]{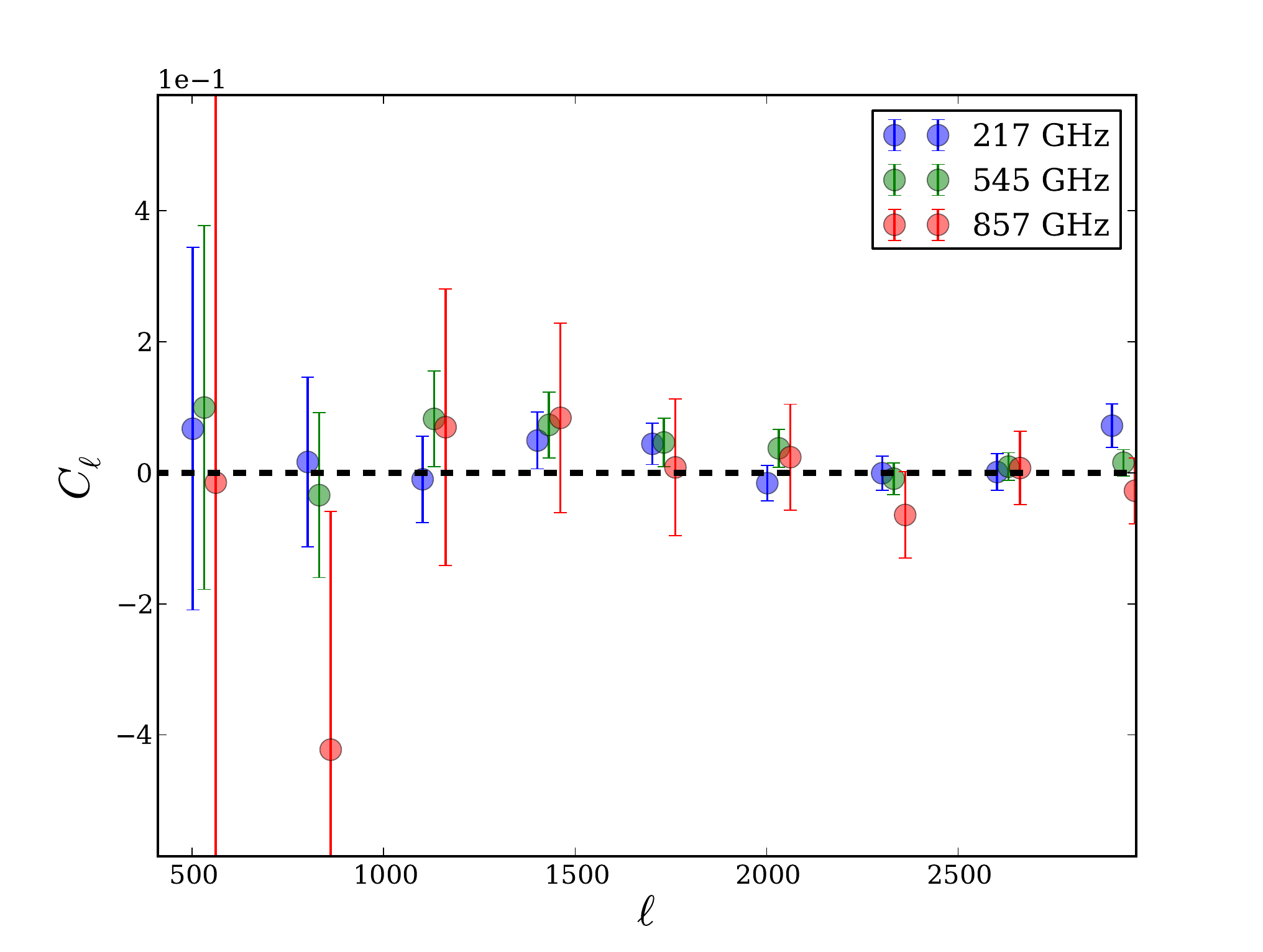}
\includegraphics[width=0.47\textwidth]{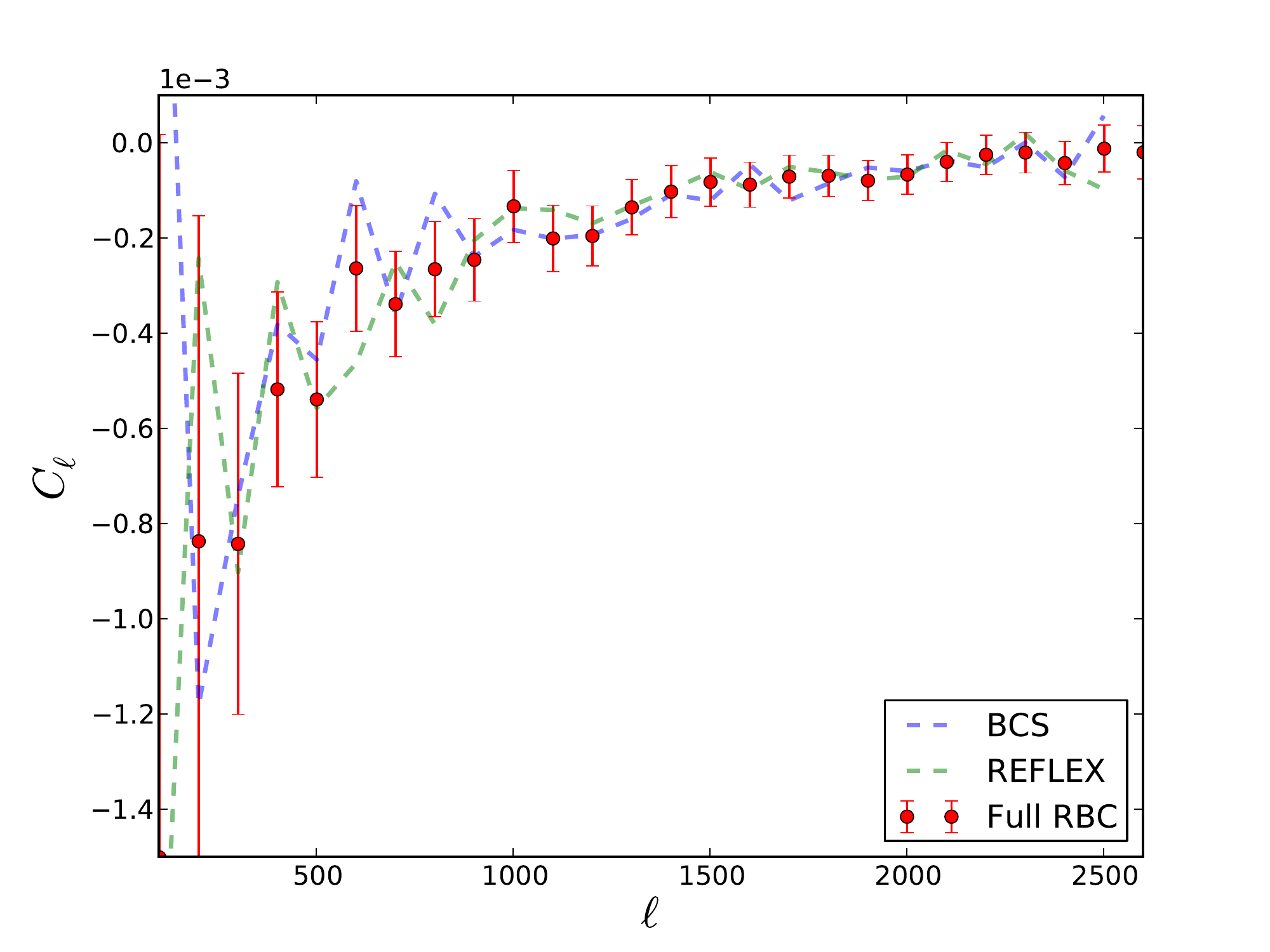}
\caption{Consistency tests: (\textit{left:}) the cross-spectra of Planck 217, 545 and 857 GHz and the cluster overdensity map used in our analysis are all consistent with zero. This constrains the CIB contamination in our results. Higher frequency maps have larger scatter around the mean due to larger galactic dust contamination and we have multiplied the 217 GHz signal by 4000 to be visible on these scales. (\textit{right:}) Cross-correlating sub-samples from REFLEX and BCS separately with the Planck 143 GHz map does not change the result. Cross-correlations of sub-samples are consistent with each other and with the main RBC sample result.}
\label{fig:tests}
\end{figure*}

The cluster sample consists of three different flux limited catalogs from three different groups using the RASS. We test the consistency and possible bias between the REFLEX and BCS samples by computing cross power spectra on the individual catalogs and comparing it to the whole RBC sample. Since there is some overlap between REFLEX and BCS we choose the celestial equator as the boundary between REFLEX (the southern sky sample) and BCS (the northern sky sample). Additionally, we exclude the 20 degrees above and below the galactic equator to exclude the CIZA sample. Right panel of Figure \ref{ts} compares the cross spectra for the REFLEX, BCS and the whole RBC sample. The uncertainty for the REFLEX and BCS sub-samples scales as $f_\rmn{sky}$ compared with the RBC errors. The $f_\rmn{sky}$ between the sub-samples and the RBC coverage is roughly 3. We find consistent results for the cross spectra of Planck with REFLEX and BCS, Fig. \ref{fig:tests}. We interpret it as an absence of a significant bias between the two catalogs, and an illustration of the consistency between the southern and northern cluster catalogs.

\section{Discussion and Conclusion}
\label{sec:concl}

\begin{figure}[t!]
\centering
\includegraphics[scale=1.0]{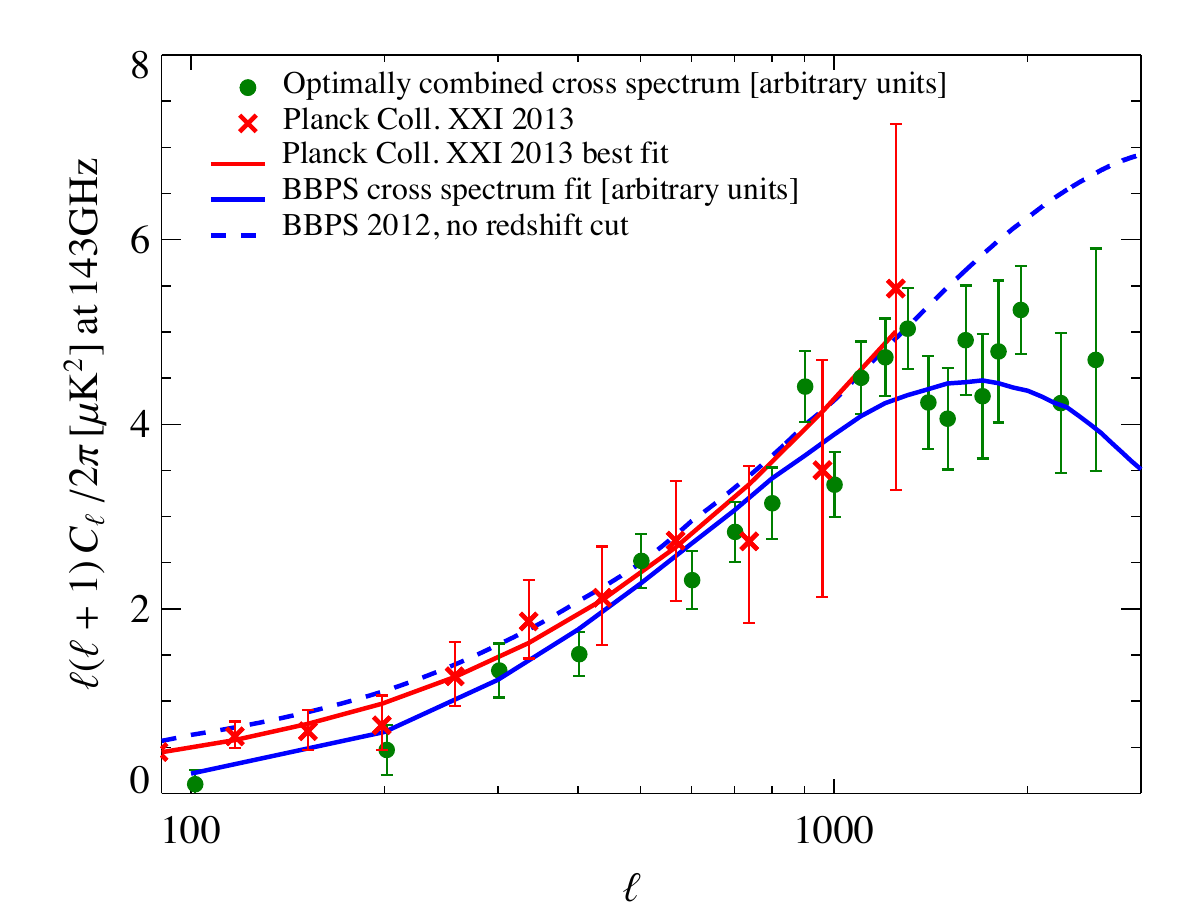}
\caption{A comparison of the Planck $y$-map auto spectrum and our optimal noise-weighted average cross spectrum. The cross spectrum (green circles) and its errors have been re-normalized to match the amplitude of the Planck $y$-map auto-power spectrum (red crosses) at $\ell=1300$ since these spectra have different units. When the clusters that contribute to these spectra are unresolved they will look Poissonian ($\ell^2C_{\ell}\propto \ell^2$). Thus, the cross spectrum starts to resolve clusters which contribute to its signal at larger angular scales than the auto spectrum because of their differences in the selection functions and form factors (see Sec. \ref{sec:ThryUn}). The tSZ auto spectrum from the same simulations used in our cross spectrum analysis (blue dashed line) agrees with the Planck auto spectrum \citep{Planck-XXI}, after accounting for the correct redshift integration.}
\label{fig:Xcomp}
\end{figure}

We present a direct measurement of the cluster power spectrum using cross-spectra of the thermal Sunyaev Zel'dovich signal in the Planck data and an overdensity map based on cluster number counts of X-ray selected clusters of galaxies in the ROSAT All Sky Survey. 
Traditionally the tSZ power spectrum is measured through fitting a multi-component model to the CMB power spectrum at frequencies below 220 GHz with the largest contribution arising from $2500 < \ell < 3500$. This results in a large variance and inevitable degeneracies with other components like the clustering of the dusty star forming galaxies in the CIB, the CIB-tSZ correlation component and the Poisson component from the radio and infrared galaxies. On the contrary the cross-correlation method is insensitive to many of the components contributing to the auto spectrum. Large components such as the CMB and most of the galactic foregrounds that do not correlate with the X-ray clusters, vanish in the two-point correlation function (i.e., the cross spectrum) and only contribute to the four-point function (i.e., the uncertainty on the cross spectrum). We construct a combined optimal cross spectrum using the 100, 143 and 343 GHz cross spectra and weighting each $\ell$ bin by the variance (see Fig \ref{fig:Xcomp}). This results in a clean and unbiased measurement of the cluster power spectrum with small uncertainties across a large range in $\ell$. We find that the error bars on this spectrum are more than 2 times smaller than the Planck $y$-map auto spectrum \citep{Planck-XXI}. The raw Planck $y$-map spectrum is contaminated by diffuse Galactic emission, clustered CIB and point source contributions. To obtain the final tSZ auto spectrum these contributions must be removed which contributes to the uncertainty. For the cross spectrum these contaminants do not correlate with cluster positions, thus do not contribute to the cross spectrum. Hence we obtain smaller error bars for the cross spectrum compared to the $y$-map auto spectrum\footnote{The error bars on the cross spectrum are similar to auto spectrum of resolved SZ sources in the Planck $y$-map analysis \citep[see Fig. 8 in ][]{Planck-XXI}}.

Our analysis is complementary to other analysis methods such as stacking while by design it will not suffer from the sub-optimality issues of stacking analysis as discussed in \citep{Siavash}. Compared to the real-space correlation function analysis, the cross-power spectrum has the same information but it enjoys the advantage of working with uncorrelated bins in the observed two-point function.

There are two cross spectra that we measured: the overdensity cross spectrum, $C_{\ell}^{SZ \times \delta_{n}}$, and the number-count cross spectrum, $C_{\ell}^{SZ \times n}$. The measurement of $C_{\ell}^{SZ \times \delta_{n}}$ is the average angular tSZ signal for the RBC cluster sample and it is sensitive to the ICM physics. The $C_{\ell}^{SZ \times n}$ measures the total angular tSZ signal for the RBC cluster sample. This signal is sensitive to both the ICM physics and cosmological parameters. The $C_{\ell}^{SZ \times n}$ signal is similar to cluster mass function measurements in that they are both measuring the exponential tail of the mass function and are highly sensitive growth of structure probes. An advantage of measuring $C_{\ell}^{SZ \times n}$ is that it weights the contribution from higher mass clusters more than lower mass clusters. Thus, it increases its sensitivity to cosmological parameters like $\sigma_8$.

Before constraining $\sigma_8$ there are theoretical modeling uncertainties associated with the cross spectrum that need to be accounted for. The cross spectrum has the advantage that it is less sensitive to {\it gastrophysics} of the ICM compared to the auto spectrum. This is because the cross spectrum only depends on one power of the Fourier transform of the cluster pressure profile, whereas the auto spectrum depends on this profile squared. The cross spectrum we calculate also requires knowledge of the cluster selection function used. The selection function has additional uncertainties associated with the masses of the clusters. Both the scatter in the $L_\rmn{X}-M$ relation and absolute mass calibration of the relation introduce biases and uncertainties into theoretical predictions. We included the scatter in the simulations through a Monte Carlo method and included a conservative estimate of $\pm 10$\% for the absolute calibration of the $L_\rmn{X}-M$. These uncertainties had a smaller effect on the cross spectrum than the ICM uncertainties when comparing the two extreme models {\it AGN feedback} and {\it shock heating}. Overall, the impact of these uncertainties are not as significant for $C_{\ell}^{SZ \times n}$ as they are for the tSZ auto spectrum and cluster number counts.

We measured a significant cross-spectrum of the clusters of galaxies (reduced $\chi^2$ = 19.1 for three non-zero spectra compared to null) using Planck tSZ data and ROSAT X-ray clusters. The {\it AGN feedback} and {\it radiative cooling} models fit the $C_{\ell}^{SZ \times \delta_{n}}$ measurements well after including a 20\% hydrostatic mass bias. We exclude at $\sim 2.5 \sigma$ level the {\it shock heating} model for the ICM and do not use it to derive cosmological constraints. We constrain the dust contamination in our analysis using the cross-power spectra of 220, 545 and 843 GHz and the RBC overdensity map. The cross spectra are shown in Figure \ref{fig:tests}. Although there seems to be a slight bump in the 545 GHz spectrum around $ 1000 < \ell < 2000 $, the errorbars are large and a significant correlation between the RBC overdensity map and the CIB can not be claimed based on these measurements. The main source of uncertainties on those scales is the galactic dust from our own galaxy. The intensity of this foreground increases with frequency because the dust spectral index is positive, causing the largest uncertainties at the highest frequency. The existence of correlation between clusters of galaxies and the CIB at significantly shorter wavelengths has been proven by cross-correlating  IRAS 100 $\mu$m and the overdensity map based on the maxBCG galaxy clusters over the cleanest region of the sky \cite{IxC}. The maps used in our analysis are at much longer wavelengths and therefore the CIB in them comes from higher redshift populations of galaxies compared to IRAS. Also as Figure \ref{fig:tests} shows, the uncertainties associated with galactic dust are large enough that a possible correlation between the RBC sample and the CIB at Planck frequencies can be neglected and absorbed into the uncertainties.

The amplitude of the $C_{\ell}^{SZ \times n}$ signal scales as $\sigma_8^{7.4}$ at $\ell \sim 1000$, thus in principle is has strong constraining power and is extremely sensitive to $\sigma_8$. Additionally this amplitude is sensitive to $\Omega_\rmn{M}$ and scales as $\Omega_\rmn{M}^{1.9}$. Without taking into account uncertainties in the astrophysical modeling of the ICM and selection function, we constrain $\sigma_8  (\Omega_\rmn{M}/0.30)^{0.26} = 0.796 \pm 0.011$ for the {\it AGN feedback} model. With the addition of WMAP9 and Planck primary constraints we obtain constraints on $\sigma_8$ of $\sigma_8 = 0.812 \pm 0.009$ and $\sigma_8 = 0.810 \pm 0.007$. After we include uncertainties in the selection function the constraint weaken to 
$\sigma_8  (\Omega_\rmn{M}/0.30)^{0.26} = 0.797 \pm 0.015$ ({\it AGN feedback} all). With the addition of WMAP9 and Planck primary constraints we obtain constraints on $\sigma_8$ of $\sigma_8 = 0.812 \pm 0.010$ and $\sigma_8 = 0.812 \pm 0.008$ (including uncertainties in the selection function). These are percent level constraints on $\sigma_8$.

How robust are these constraints given the ICM models used in the simulations? First, we compared the simulations against the average spectrum $C_{\ell}^{SZ \times \delta_{n}}$ and then derived $\sigma_8$ values from the total spectrum $C_{\ell}^{SZ \times n}$. If the ICM model in the simulations disagreed with the cluster sample this disagreement would appear when comparing the $C_{\ell}^{SZ \times \delta_{n}}$ signals. Both the $C_{\ell}^{SZ \times \delta_{n}}$ from the {\it AGN feedback} and {\it radiative cooling} models agree with the measured $C_{\ell}^{SZ \times \delta_{n}}$, while the {\it shock heating} model was $\sim 2.5 \sigma$ different. Second, the cross spectrum is less sensitive to a given ICM model since it is $\propto \tilde{y}$ and not $\propto |\tilde{y}|^2$ like the auto spectrum where large differences between the auto spectra amplitudes of these models are found \citep{Batt2010}. Third, stacked measurements of the cluster pressure profiles \citep[e.g.,][]{arnd2010,PlanckV,Saye2013} are in agreement with the {\it AGN feedback} model. Even the extreme {\it shock heating} model, which is already in disagreement with most cluster observations, has a derived $\sigma_8$ value that is within $1 \sigma$ of the {\it AGN feedback}. Thus, the derived $\sigma_8$ parameters from {\it AGN feedback} models (and {\it radiative cooling} model) are robust.

Currently the $\sigma_8$-only constraints in this analysis are obtained after including primary CMB constraints from WMAP and Planck. With the imminent all sky X-ray survey from eRosita, this will not be necessary. The expectation is that eRosita will provide a larger flux limited all sky cluster survey to higher redshifts. Thus, one would have a larger sample of clusters with a higher signal to noise ratio such that the cross spectrum could be divided up into redshift slices and break the $\sigma_8$ and $\Omega_\rmn{M}$ degeneracy since the parameter degeneracy directions in the $\sigma_8-\Omega_\rmn{M}$ plane at different redshifts are slightly different \citep[e.g.,][]{satej2010}. Then, in order to fully exploit this cross spectrum measurement one would require a better understanding of the cluster selection function, more specifically a higher precision mass calibration for the $L_\rmn{X} - M$ scaling relation would be needed. Joint SZ, X-ray and optical observations of nearby clusters should improve the cluster mass calibrations and constrain the ICM {\it astrophysics}. Additionally, the next Planck data release will have higher fidelity maps and smaller errors on $\sigma_8$ and $\Omega_\rmn{M}$. Thus, the constraint on $\sigma_8$ using $C_{\ell}^{SZ \times n}$ will become tighter.

Constraining $\sigma_8$ with the cross correlation of tSZ and X-ray cluster catalogs is just a first step. In principle, this method can be used to constrain other cosmological parameters that are sensitive to growth of structure measurements (similar to the cluster mass function) such as the sum of neutrino masses, $\Sigma\nu_\rmn{e}$, or the dark energy equation of state, $w$.

\bigskip

We would like to thank K. Huffenberger for guidance on using the 
appropriate beam information in the Planck data release. We enjoyed discussions with Jack Hughes and Felipe Menanteau in the early stages of this work and would like to thank them. We also acknowledge useful discussions with Matt Dobbs, Colin Hill, Gil Holder, Eiichiro Komatsu, Etienne Pointecouteau, Hy Trac and Keith Vanderlinde. Some of the results in this paper have been derived using \textit{healpy}, the python version of the HEALPix package \citep{healpix}. N.B. is supported by a McWilliams Center for Cosmology Postdoctoral Fellowship made possible by Bruce and Astrid McWilliams Center for Cosmology. C.P. gratefully acknowledges financial support of the Klaus Tschira Foundation.  DNS acknowledges support from the NASA Theory program grant NNX12AG72G.  Research in Canada is supported by NSERC and CIFAR. Simulations and computations were performed on the GPC supercomputer at the SciNet HPC Consortium. SciNet is funded by the CFI under the auspices of Compute Canada, the Government of Ontario, the Ontario Research Fund Research Excellence; and the University of Toronto.

\bibliographystyle{plain}

\end{document}